\documentclass[12pt,a4paper]{article}

\usepackage{a4wide}
\usepackage{amsmath,amssymb}
\usepackage{bm}
\usepackage{amssymb}
\usepackage{hyperref}
\usepackage{epic}

\newcommand{\LoverlogM}{j}
\newcommand{\holegap}{c}
\newcommand{\magnonrho}{\rho_{\rm m}}
\newcommand{\magnonbarrho}{\bar \rho_{\rm m}}
\newcommand{\korrho}{\rho_0}
\newcommand{\korbarrho}{\bar \rho_0}
\newcommand{\holerho}{\rho_{\rm h}}
\newcommand{\holebarrho}{\bar \rho_{\rm h}}
\newcommand{\oldtildesigma}{\hat{\sigma}}

\newcommand{\pint}{\makebox[0pt][l]{\hspace{3.1pt}$-$}\int}

\newcommand{\Tr}{{\rm Tr \,}}
\newcommand{\Op}{\mathcal{O}}
\newcommand{\fldZ}{\mathcal{Z}}

\newcommand{\fldD}{\mathcal{D}}

\newcommand{\alg}[1]{\mathfrak{#1}}
\newcommand{\superN}{\mathcal{N}}

\newcommand{\indups}[1]{_{\mathrm{\scriptscriptstyle #1}}}
\newcommand{\gym}{g\indups{YM}}

\newcommand{\sfrac}[2]{{\textstyle\frac{#1}{#2}}}
\newcommand{\half}{\sfrac{1}{2}}

\makeatletter
\def\mr@ignsp#1 {\ifx\:#1\@empty\else #1\expandafter\mr@ignsp\fi}%
\newcommand{\multiref}[1]{\begingroup
\xdef\mr@no@sparg{\expandafter\mr@ignsp#1 \: }%
\def\mr@comma{}%
\@for\mr@refs:=\mr@no@sparg\do{\mr@comma\def\mr@comma{,}\ref{\mr@refs}}%
\endgroup}
\makeatother

\numberwithin{equation}{section}

\makeatletter
\let\old@startsection=\@startsection
\renewcommand{\@startsection}[6]{\old@startsection{#1}{#2}{#3}{#4}{#5}{#6\mathversion{bold}}}
\makeatother

\begin{document}
\thispagestyle{empty}

\begin{flushright}\footnotesize
\texttt{AEI-2007-173}\\
\texttt{UUITP-19/07}\\
\texttt{NI07091}\\
\vspace{0.5cm}
\end{flushright}
\setcounter{footnote}{0}

\begin{center}
{\Large{\bf A Generalized Scaling Function for AdS/CFT }}
\vspace{15mm}

{\sc Lisa Freyhult $^{a,b}$, Adam Rej $^a$ and 
Matthias Staudacher $^{a,c}$}\\[5mm]

{\it $^a$ Max-Planck-Institut f\"ur Gravitationsphysik\\
    Albert-Einstein-Institut \\
    Am M\"uhlenberg 1, D-14476 Potsdam, Germany}\\[5mm]

{\it $^b$ Department of Theoretical Physics\\
    Uppsala University \\
    SE-751 08 Uppsala, Sweden}\\[5mm]

{\it $^c$ Isaac Newton Institute for Mathematical Sciences\\
    20 Clarkson Road, Cambridge, CB3 0EH, U.K. }\\[5mm]

\texttt{lisa.freyhult@aei.mpg.de}\\
\texttt{arej@aei.mpg.de}\\
\texttt{matthias@aei.mpg.de}\\[30mm]

\textbf{Abstract}\\[2mm]
\end{center}

\noindent{We study a refined large spin limit for twist operators in the
$\alg{sl}(2)$ sector of AdS/CFT. We derive a novel non-perturbative 
equation for the generalized two-parameter scaling function 
associated to this limit, and analyze it at weak coupling. It is 
expected to smoothly interpolate between weakly  coupled gauge theory and string theory at strong coupling.
}
\newpage

\setcounter{page}{1}
\section{Introduction, Main Result, and Open Problem}
\label{sec:intro}

The perhaps most interesting subset of all local composite quantum
operators of planar $\superN=4$ supersymmetric gauge theory is 
formed by the sector of $\alg{sl}(2)$ twist operators. 
The reason is that these bear many similarities with the 
twist operators of QCD. They may be symbolically written as
\begin{equation}\label{ops}
\Tr \big( \fldD^M
\fldZ^L \big)+\ldots \, ,
\end{equation}
which is a shorthand notation for intricate linear superpositions of
all states where the $M$ covariant derivatives $\fldD$
act in all possible ways on the $L$ complex scalar fields $\fldZ$.
Here $L$ is a $\alg{su}(4)$ R-charge, frequently denoted as
$J$ in the literature, and $M$ is a Lorentz spin, often called $S$.
Our labelling refers to the magnetic spin chain picture of
these operators, where $L$ is the length of the chain, and
$M$ is the ``magnon number''. The twist of an operator is defined
as the classical dimension minus its Lorentz spin, so 
the length $L$ equals the twist in the case of \eqref{ops}.

$\superN=4$ gauge theory is a superconformal field theory. Therefore
proper superpositions of the operators \eqref{ops} must carry
a definite charge $\Delta$ under dilatations. It generically is, in 
contradistinction to the R-charge $L$ and the Lorentz charge $M$, coupling constant dependent: $\Delta=\Delta(g)$. Its anomalous part
$\gamma(g)$ is defined as\footnote{%
The anomalous dimension $\gamma(g)$ is related to
the energy $E(g)$ of the integrable long range spin chain describing
the operators \eqref{ops} through $\gamma(g)=2\,g^2\,E(g)$.
It should not be confused with the energy of string states
which equals $\Delta(g)$ via the AdS/CFT correspondence.
}%
\begin{equation}\label{spinenergy}
\Delta(g)=M+L+\gamma(g)\, ,
\end{equation}
where $M+L$ is the classical dimension of the operators \eqref{ops}. 
In the case of the operators \eqref{ops} $\gamma(g)$ behaves in a
very interesting way as the spin $M$ gets large at fixed
twist $L$. It grows logarithmically with $M$ at all orders of
the coupling constant $g$ defined as
\begin{equation}\label{convention}
g^2 \, = \, \frac{\gym^2 \, N}{8 \, \pi^2} =
\frac{\lambda}{16 \, \pi^2}\, ,
\end{equation}
where $\lambda$ is the 't Hooft coupling.
The prefactor of the logarithm is a function of $g$. We call it the
{\it universal scaling function} $f(g)$:
\begin{equation}\label{log}
\Delta -M -L  = \gamma(g)=f(g)\; \log M +\ldots \, .
\end{equation}
This behavior is a special case of so-called Sudakov scaling,
see \cite{Collins:1989bt}. In the twist $L=2$ case it equals {\it twice}
the cusp anomalous dimension of light-like Wilson loops
\cite{Korchemsky:1988si}. The independence or ``universality''
of the function $f(g)$ on the twist $L$, with $L$ arbitrary but finite,
or even $L \rightarrow \infty$ as long as $L \ll \log M$,  
was first pointed out at one loop in \cite{Belitsky:2006en}, 
and conjectured to hold at arbitrary loop order in \cite{Eden:2006rx}.
It would be very interesting to rigorously prove that the
twist-two ($L=2, M \rightarrow \infty$), 
twist-$L$ ($L$ fixed, $M \rightarrow \infty$),
and the universal ($L,M \rightarrow \infty, L \ll \log M$)
scaling functions $f(g)$ of the operators \eqref{ops} indeed all
coincide for arbitrary values of $g$: 
$f(g)=f^{(2)}(g)=f^{(3)}(g)=\dots=f^{(L)}(g)=\ldots=f^{(\infty)}(g)$.

Based on the conjectured all-loop integrability of planar $\superN=4$ 
theory \cite{Beisert:2003tq}, the weak coupling expansion of $f(g)$
is known from the solution of an integral equation obtained from
the asymptotic Bethe ansatz for these operators
\cite{Beisert:2005fw,Eden:2006rx,Beisert:2006ez}.
It agrees to four orders\footnote{To be precise, the four-loop 
field-theory result of \cite{Bern:2006ew} agrees numerically with the analytic Bethe ansatz prediction in \eqref{weakBES} to 0.001\%. 
An analytic proof would be most welcome.}
with field theory \cite{Bern:2006ew}:
\begin{equation}\label{weakBES}
f(g)=8\, g^2
-\frac{8}{3}\, \pi^2 g^4
+\frac{88}{45}\, \pi^4  g^6
- 16\left(\frac{73}{630}\,\pi^6 + 4\, \zeta(3)^2 \right)g^8
\pm\ldots\, .
\end{equation}
Testing the Bethe ansatz to five orders in field theory might
not be out of reach \cite{Bern:2007ct}.

The strong coupling expansion may also be obtained from
the same integral equation \cite{Beisert:2006ez} which generates
the small $g$ expansion \eqref{weakBES}. After the
initial studies \cite{Kotikov:2006ts,Benna:2006nd},
an impressive analytical expansion method to any desired order
was worked out in \cite{Basso:2007wd}.
The starting point of this systematic approach was an important
decoupling method discovered by Eden \cite{Eden},
see also \cite{Kotikov:2006ts}. The series starts as
\begin{equation}\label{strongBES}
f(g)=4\, g
-\frac{3\,\log 2}{\pi}
-\frac{\rm{K}}{4\,\pi^2} \frac{1}{g}
-\ldots\, ,
\end{equation}
where K$=\beta(2)$ = Catalan's constant.
The first two terms on the r.h.s.~agree, respectively, with the
classical and one-loop \cite{Gubser:2002tv,Frolov:2002av} result
from semi-classical string theory, and the third term
is the two-loop correction very recently obtained in
\cite{Roiban:2007jf,Roiban:2007dq}.
It would be very interesting to also check the three-loop term
in string theory\footnote{%
However, one important fact to keep in mind is that
the weak coupling expansion \eqref{weakBES} has a
finite radius of convergence, while the strong coupling series
\eqref{strongBES} is asymptotic, and, apparently, not even
Borel-summable \cite{Basso:2007wd}.
So \eqref{strongBES} follows from knowing all terms in
\eqref{weakBES}, but, conversely, knowing all terms
of the string expansion \eqref{strongBES} does not
allow to reconstruct the gauge-theoretic perturbation series
\eqref{weakBES} without further input. Unfortunately, it
is currently not even known what the nature of this input might be.
}.%

So it appears that $f(g)$ is the first example of an
exactly known, via the solution of a linear integral equation
\cite{Beisert:2006ez}, function which smoothly {\it interpolates} between a gauge theory and a string theory observable in the
AdS/CFT system. A natural question is whether further
interesting examples may be found, and whether the function
$f(g)$ may be generalized. A major obstacle is the fact
that we currently only know the {\it asymptotic} spectrum of
the planar $\superN=4$ model, as was recently unequivocally
established in \cite{Kotikov:2007cy}. Important clues come
from both taking a closer look at the scaling law \eqref{log} in
the one-loop gauge theory \cite{Belitsky:2006en}, and
at intriguing string theory results \cite{Frolov:2006qe,Alday:2007mf} 
generalizing the expansion \eqref{strongBES}. Put together,
these suggest that at weak coupling an interesting  
generalized scaling limit might exist, where\footnote{The
variable $j$ was first explicitly introduced (up to a factor of $1/2$)
in eq.~(3.1) of \cite{Alday:2007mf}.
}%
\begin{equation}\label{newlimit}
M \rightarrow \infty\, ,
L \rightarrow \infty\, ,
\qquad {\rm with} \qquad
\LoverlogM :=  \frac{L}{\log M}={\rm fixed}\, .
\end{equation}
%
We will prove in this paper that this is indeed the
case, first at one-loop order, and then beyond. 
More precisely, we will show that a {\it generalized} scaling function
$f(g,  \LoverlogM)$ exists to all orders in perturbation theory
\begin{equation}\label{newlog}
\Delta -M -L = \gamma(g)=f(g,  \LoverlogM) \log M
+\ldots \, ,
\end{equation}
where $f(g,0)=f(g)$ in \eqref{log}.
This extends the one-loop results in \cite{Belitsky:2006en},
and the all-loop result at $\LoverlogM=0$ of
\cite{Eden:2006rx,Beisert:2006ez}.
The latter is possible since in the limit \eqref{newlimit}
$L \rightarrow \infty$, and we may therefore use the asymptotic Bethe
ansatz methodology of 
\cite{Beisert:2005fw,Eden:2006rx,Beisert:2006ez}.

The final result of our analysis, presented in detail in the 
ensuing chapters, is the following integral equation
\begin{equation}\label{besartige}
\oldtildesigma(t)=\frac{t}{e^t-1}\left(\hat {\cal K}(t,0)-4\,\int^{\infty}_{0}dt'\,\hat {\cal K}(t,t')\,\oldtildesigma(t')\right).
\end{equation}
It is essentially identical in form to the ``ES'' (no dressing phase)
\cite{Eden:2006rx} and ``BES'' (with proper dressing phase) 
\cite{Beisert:2006ez} equations.
The kernel corresponding to the generalized scaling limit is
quite involved as it contains various contributions. It reads
\begin{eqnarray}\label{kern}
\hat {\cal K}(t,t')&=&g^2\,\hat K(2gt,2gt')+\hat{K}_{\rm h}(t,t';a)-\frac{J_0(2gt)}{t}\,\frac{\sin{at'}}{2\,\pi\,t'}\,e^{\frac{t'}{2}}\nonumber\\
&-&4\,g^2\,\int_0^\infty dt''\,t''\,\hat K(2gt,2gt'')\,\hat{K}_{\rm h}(t'',t';a).
\end{eqnarray}
Here
\begin{equation}\label{beskern}
\hat{K}(t,t')=\hat{K}_0(t,t')+\hat{K}_1(t,t')+\hat{K}_{\rm d}(t,t')
\end{equation}
is the kernel of the ``BES'' equation, where 
\begin{equation}\label{knull}
\hat{K}_0(t,t')=\frac{t\,J_1(t)\,J_0(t')-t'\,J_0(t)\,J_1(t')}{t^2-t'^2}\, ,
\end{equation}
\begin{equation}\label{keins}
\hat{K}_1(t,t')=\frac{t'\,J_1(t)\,J_0(t')-t\,J_0(t)\,J_1(t')}{t^2-t'^2}\, ,
\end{equation}
and the kernel encoding the effects of the dressing phase is given 
by the convolution
\begin{equation}\label{kdressing}
\hat{K}_{\rm d}(t,t')=8\,g^2\,\int^{\infty}_{0} dt''\, \hat{K}_1(t,2gt'')\,\frac{t''}{e^{t''}-1}\,\hat{K}_0(2gt'',t')\, .
\end{equation}
For a possible mechanism generating this type of convolution
structure see \cite{Rej:2007vm}.
The novel contributions generated by a non-vanishing $\LoverlogM$
are encoded in the kernel
\begin{equation}\label{ktilde}
\hat{K}_{\rm h}(t,t';a)=\frac{1}{2\,\pi\,t}\,e^{-\frac{t}{2}}\,\frac{t\cos(at')\sin(at)-t'\cos(at)\sin(at')}{t^2-{t'}^2}\,e^{\frac{t'}{2}}\, ,
\end{equation}
as well as the explicit, rightmost term of the first line of \eqref{kern},
and the further convolution in the second line of that equation.
The index h of $\hat{K}_{\rm h}(t,t';a)$ stands for ``hole'', its meaning
will become clear below. The corrections of the refined limit depend
on a ``gap'' parameter $a$ whose interpretation will also be explained.
Its relation to $j$ is fixed by the constraint
\begin{equation}\label{normalizationtspace}
\LoverlogM=\frac{4\,a}{\pi}-\frac{16}{\pi}\int^{\infty}_{0} dt \,\hat{\sigma}(t)\, e^{\frac{t}{2}}\, \frac{\sin{at}}{t}\, .
\end{equation}
Lastly, the generalized scaling function of \eqref{newlog} is
given by
%
%
%
\begin{equation}
\label{genscalingfunction}
f(g,\LoverlogM)=
\LoverlogM+16\,\oldtildesigma(0)\, .
\end{equation}
It is determined by first solving the integral equation \eqref{besartige}
with the kernel \eqref{kern} for the fluctuation density 
$\oldtildesigma(t)=\oldtildesigma(t;g,a)$ as a function of 
$g$ and $a$. Then $a$ is
found as a function of $j$ by inverting the relation 
\eqref{normalizationtspace}, i.e.~by computing $a(\LoverlogM)$. 
This then yields $\oldtildesigma(t)=\oldtildesigma(t;g,a(\LoverlogM))$ 
as a function of $g$ and $\LoverlogM$, and the generalized 
scaling function $f(g,\LoverlogM)$ is finally obtained by evaluating 
the latter at  $t=0$, see  \eqref{genscalingfunction}. 

As in \cite{Eden:2006rx,Beisert:2006ez}, in practise it appears
impossible to produce a closed-form solution of the equation
\eqref{besartige}. In fact, we did not even find an explicit solution
at one-loop order, i.e.~for $g=0$. It is however possible to solve it 
iteratively in a double-expansion in small $g$ and small $j$. Excitingly,
the obtained function appears to be ``bi-analytic'', i.e.~analytic in
$g$ around $g=0$ at arbitrary finite values of $\LoverlogM$, and vice versa. We therefore believe that our equations actually hold for 
{\it arbitrary} values of $g$ and $j$. The beginning of this double
expansion may be found in \eqref{f1},\eqref{f2},\eqref{f3},\eqref{f4},
which we have displayed by giving the four-loop result of the
functions $f_1(g) \ldots f_4(g)$ defined through
\begin{eqnarray}\label{f-family}
f(g,\LoverlogM)=f(g)+\sum_{n=1}^\infty f_n(g)\,\LoverlogM^n\, .
\end{eqnarray}
Our truncation at four loops $\Op(g^8)$ and $\Op(\LoverlogM^4)$
is due to space limitations, and one easily generates many
more orders in $g^2$ and $\LoverlogM$ if needed.
A curious fact is the absence of any terms of $\Op(j^2)$,
i.e.~the function $f_2(g)$ is zero. We will come back to this
point shortly.

A very interesting question is how $f(g,  \LoverlogM)$
behaves at strong coupling. Indeed we would like to make
contact with the already known results from string theory
\cite{Frolov:2002av,Beisert:2003ea,Belitsky:2006en,
Frolov:2006qe,Alday:2007mf,Radu+Arkady}.
A potential trouble is that
in the semi-classical computations pioneered by
Frolov and Tseytlin \cite{Frolov:2002av} the coupling constant $g$ is
intricately entangled with the, respectively, $AdS_5$ and $S^5$
charges $M$ and $L$. In \cite{Belitsky:2006en} the 
strong coupling limit of the dimension $\Delta$ of the operators 
\eqref{ops} was predicted from the results of \cite{Frolov:2002av,Beisert:2003ea} on the energy of a folded string soliton
(see also the discussion in 
\cite{Frolov:2006qe,Alday:2007mf,Radu+Arkady,Casteill:2007ct}).
The prediction reads
\begin{equation}\label{FT}
\Delta_{\rm classical}=M+L\,\sqrt{1+z^2}\, + \ldots\, ,
\end{equation}
where\footnote{The contemporaneously appearing
work \cite{Radu+Arkady} uses the notation
$\ell=1/z$ and $\Lambda=4\,\pi\,g$.
}%
\begin{equation}\label{FTconds}
M \rightarrow \infty\, ,
L \rightarrow \infty\, ,
g \rightarrow \infty\, ,
\quad {\rm with} \quad
M \gg L
\quad {\rm and~fixing}\quad
\frac{M}{g}\, ,
\frac{L}{g}\, ,
z:=4\,g\,\frac{
\log \frac{M}{\Lambda} }
{L}\, ,
\end{equation}
and $\Lambda$ is some scale\footnote{%
The scale $\Lambda$ actually being 
used in the string theory calculations in
\cite{Frolov:2002av,Beisert:2003ea,Belitsky:2006en,
Frolov:2006qe,Alday:2007mf,Radu+Arkady} seems to be
somewhat unclear.
Is the proper scale (1) $\Lambda=4\,\pi\,g$  or (2) $\Lambda=L$
or (3) $\Lambda=1$? Since these calculations start from fixing
$M/g$ and $L/g$ it would seem that they require either (1) or (2).
At weak coupling we definitely have (3), as we are proving
to all orders in this paper. 
Understanding the crossover of scales as one moves from
weak to strong coupling, or vice versa, should be very interesting.
}. %
The result \eqref{FT} was derived in \cite{Casteill:2007ct} from
the asymptotic Bethe ansatz
\cite{Beisert:2005fw,Beisert:2006ez} with approximate
strong coupling (AFS \cite{Arutyunov:2004vx} only) 
dressing phase. While this constitutes an important check of 
the Bethe ansatz method, this had to work in the sense that the 
dressing phase \cite{Arutyunov:2004vx} was extracted from
the integrable structure of classical string theory
\cite{Bena:2003wd,Kazakov:2004qf}. What is
missing is a derivation from a solution of the exact
ansatz \cite{Beisert:2005fw,Beisert:2006ez}
which interpolates between weak and strong coupling.
Now it is tempting to identify, in view of \eqref{newlimit},
\begin{equation}\label{identify}
z=\frac{4\,g}{\LoverlogM}\, .
\end{equation}
Certainly the condition $M \gg L$ with $M,L \rightarrow \infty$ is
satisfied in the weak-coupling limiting procedure \eqref{newlimit}.
The more questionable assumptions in semi-classical string theory,
as far as concerns extrapolating weak coupling results,
are the fixation of $M/g$, $L/g$ (see also footnotes on the
previous page). Proceeding under this caveat
we could then rewrite \eqref{FT} as
\begin{equation}\label{FTrewritten}
\Delta_{\rm classical}-M-L=
\Bigg(4\,g\,\sqrt{1+\left(\frac{\LoverlogM}{4 g}\right)^2}
-\LoverlogM \Bigg)
\, \log M\,
+ \ldots\, .
\end{equation}
If we now also expand in small $\LoverlogM$ we find
\begin{equation}\label{FTexpanded}
\Delta_{\rm classical}-M-L=
\Bigg(4\,g-\LoverlogM
+\frac{\LoverlogM^2}{8\,g}+\Op(\LoverlogM^4)\Bigg)
\, \log M\,
+ \ldots\, .
\end{equation}
The leading term $4 g$ agrees with the first term in
\eqref{strongBES}.
The one-loop string correction to \eqref{FT} was computed in
\cite{Frolov:2006qe}
\begin{eqnarray}\label{FTT}
\lefteqn{\Delta_{\rm 1-loop}
=\frac{L}{\sqrt{\lambda}}
\frac{1}{\sqrt{1+z^2}} \left\{ z\sqrt{1+z^2}-(1+2z^2)
\log\left[z+\sqrt{1+z^2}\right]\right. }\nonumber \\
&&\hspace{1.5cm}\left.-z^2+2(1+z^2)\log(1+z^2)
-(1+2z^2)\log\left[\sqrt{1+2z^2}\right]
\right\}. \label{E1}
\end{eqnarray}
Taking $z\rightarrow \infty$ it produces the second term
on the r.h.s.~of \eqref{strongBES}. If we were to again expand in
small $\LoverlogM$ via \eqref{identify} we would find
$\LoverlogM^2 \log \LoverlogM$ terms.
The result \eqref{FTT} was fully derived in \cite{Casteill:2007ct} from
the asymptotic Bethe ansatz
\cite{Beisert:2006ez,Beisert:2005fw} with approximate
strong coupling (AFS + HL \cite{Hernandez:2006tk}) dressing phase.
Once again, this is an important cross-check on the consistency of
the extraction of the one-loop correction of the dressing phase
from one-loop string theory \cite{Hernandez:2006tk,Freyhult:2006vr}, 
see also the very recent derivation \cite{Gromov:2007ky},
but does not answer the question how the dimensions of the
gauge theory states in \eqref{ops} ``flow'' to the energies of
string theory states as the coupling increases.

An interesting insight into the structure of further 
quantum corrections, i.e.~two-loop and higher, 
to \eqref{FT},\eqref{FTT}  was obtained in \cite{Alday:2007mf}
in the limit $z \rightarrow \infty$. In
a paper contemporaneous with ours \cite{Radu+Arkady},
an impressive direct two-loop string calculation, in this limit,
is performed which agrees with the results of \cite{Alday:2007mf}.
However, Roiban and Tseytlin argue in \cite{Radu+Arkady}
that after resumming infinitely many terms of the form 
$\LoverlogM^2\, \log^k \LoverlogM$ all terms of the
form $\LoverlogM^2$ might vanish. They furthermore
noticed that some initial support for these considerations is
provided by a fascinating and curious byproduct of 
our derivation: The function $f_2(g)$ in the expansion 
\eqref{f-family} is exactly zero! 
This suggests that extrapolation between the result at small 
$g$ and the result at large $g$ might indeed work out.

Therefore an exciting open problem not addressed in this paper 
is to now solve our equations
at strong coupling $g \rightarrow \infty$ in order to see
whether any of the above string results are reproduced,
and whether the extrapolation works out.
In fact, our derivation does not assume $j$ to be small,
so we are hopeful that under the identification \eqref{identify}
the full strong coupling expansion of $f(g,\LoverlogM)$, 
i.e.~\eqref{FT},\eqref{FTT} and
all further corrections, in generalization of the beautiful
expansion of \cite{Basso:2007wd} at $\LoverlogM=0$, 
will be obtained. As already mentioned this is however
not assured, as we might run into an order-of-limits problem,
namely \eqref{newlimit} versus \eqref{FTconds}.
It would also be important to gain an understanding how the
states corresponding to the generalized scaling function
fit into the general classification of classical integrable
curves \cite{Kazakov:2004qf,Dorey:2006zj} and their
quantum fluctuations. It should be very interesting
to see how the parameter $\LoverlogM$
in \eqref{newlimit} relates to the ``filling fractions'' of the
classical curve.

This paper is organized as follows. In section \ref{sec:1-loop}
we extend the study in \cite{Belitsky:2006en} and take a close 
look at the fine-structure of the large-spin $M$ 
anomalous dimensions of \eqref{ops} 
at one-loop order. We derive our results both using
traditional techniques as well as more sophisticated ones
involving so-called non-linear integral (or also
``Destri-DeVega'') equations, see \cite{Feverati:2006tg} and references therein. In section \ref{sec:all-loop} we
generalize the methodology of the 
non-linear integral equations to all orders in the coupling
constant and compute some novel finite size $\Op(M^0)$ corrections
to the scaling behavior \eqref{log}. In section \ref{sec:genscaling}
we extend our one-loop results to all loops, prove
the existence of the novel generalized scaling function in 
\eqref{newlog}, and derive the above equations determining it.

\section{One-Loop Theory}
\label{sec:1-loop}

\subsection{Magnons and Holes}
\label{sec:mandh}

The one-loop diagonalization problem of the operators \eqref{ops}
is equivalent to the one of an integrable spin chain with $\alg{sl}(2)$
symmetry. This was first discovered in 
\cite{Lipatov:1997vu,Braun:1998id} and more specifically in the
$\superN=4$ context,
extending the discovery of \cite{Minahan:2002ve},
in \cite{Beisert:2003yb}. The allows to apply the Bethe
ansatz, which then leads to the following
one-loop Bethe equations 
\begin{equation}\label{sl2}
\bigg(\frac{u_k+\frac{i}{2}}{u_k-\frac{i}{2}}\bigg)^L =
\prod^{M}_{\substack{j=1 \\j\neq k}}\frac{u_k-u_j-i}{u_k-u_j+i}\, ,
\end{equation}
where $L$ is the length (=twist in this case)
and $M$ is the number of magnons, see \eqref{ops}.
The cyclicity constraint and the one-loop
anomalous dimension $\gamma_1$
(see \eqref{spinenergy}) are
\begin{equation}\label{1loopmomeng}
\prod_{k=1}^M\,\frac{u_k+\frac{i}{2}}{u_k-\frac{i}{2}}=1
\qquad {\rm and} \qquad
\gamma_1=\frac{\gamma(g)}{g^2} \bigg\vert_{g=0}=
2\,\sum_{k=1}^M\, \frac{1}{u_k^2+\frac{1}{4}}\, .
\end{equation}
With the help of the Baxter function
\begin{equation}
Q(u)=\prod_{k=1} ^{M} (u-u_k)
\end{equation}
one can write down an off-shell version of these equations
\begin{equation}\label{bax}
\left(u+\frac{i}{2} \right)^L \,Q(u+i)+\left(u-\frac{i}{2} \right)^L \,Q(u-i)\,=\,t(u)\, Q(u)\, ,
\end{equation}
where
\begin{equation}\label{trm}
t(u)= 2\,u^L+\sum_{i=2}^L q_i\, u^{L-i}
\end{equation}
is the transfer matrix given in terms of the charges. The ground state for arbitrary $L$ and $M$ is unique and thus the corresponding charges are fixed.
Clearly setting $u=u_k$ in equation \eqref{bax} brings us back to \eqref{sl2}. However one of the advantages of \eqref{bax} is the possibility of identification of complementary solutions
$u=u_{\rm h}^{(k)}$ to \eqref{sl2} \cite{Belitsky:2006en}. They are 
found as the zeros of the transfer matrix, i.e.~from $t(u)=0$
and describe ``holes''. We thus have
\begin{equation}\label{holes}
t(u)=2\,\prod_{k=1}^L(u-u_{\rm h}^{(k)})\, .
\end{equation}
We can intuitively think of the
hole roots as rapidities describing the motion of the
$\fldZ$-particles in the spin chain interpretation of the
operators \eqref{ops}.
For a general value of $L$ the equation \eqref{holes} has $L$ solutions and thus there are $L$ holes. One can prove
that for any state all magnon roots $u_k$ and all hole
roots $u_{\rm h}^{(k)}$ are real.
It is possible to find  the $q_2$ charge analytically by matching the three highest powers
of $u$ in the Baxter equation \eqref{bax}:
\begin{equation}\label{q2}
q_{2}\, =\, -\frac{1}{4}\,L\,(L-1)-L\,M-M\,(M-1) \, .
\end{equation}
Because $q_{2}$ and all higher charges explicitly depend on $M$ the roots of $t(u)$ will also, generically, depend on $M$. One can
argue, however, that for the groundstate, and in the case
$L \ll M$, two of them
are special, see \cite{Belitsky:2006en}: Their magnitude
is larger than the one of any other (hole or magnon) Bethe root,
and scales with $M$ as the magnon number $M$ gets large,
see \cite{Belitsky:2006en}.
To identify these roots one recalls that the mode numbers for
magnons for the ground states, when $L \ll M$, 
are given by \cite{Eden:2006rx}
\begin{equation}\label{modenumbers}
n_k=k+\frac{L-3}{2}\, \mbox{sgn}(k) \qquad \mbox{for} \qquad k \,=\,\pm 1 \pm 2,...,\pm \frac{M}{2}\, .
\end{equation}
The absolute value of the roots grows monotonically with $|n_k|$. It
follows from \eqref{modenumbers} that the rapidities of the magnons
and holes are parity-invariant. Among the holes there are two
'universal holes' which occupy the highest allowed mode numbers
\begin{equation}\label{modenumbersuniversalholes}
n^{u,1}_{\rm h}=\frac{L+M-1}{2} \qquad n^{u,2}_{\rm h}=-\frac{L+M-1}{2}\, .
\end{equation}
The corresponding hole roots are precisely the one that scale with $M$.
The remaining holes fill the gap in the mode numbers of magnons
\begin{equation}\label{modenumbersholes}
n^{r}_{\rm h} \in \left\{-\frac{L-3}{2}, \ldots, \frac{L-3}{2} \right\}.
\end{equation}
For the ground state, when $L \ll M$, the
magnitudes of the roots are thus ordered as
\begin{equation}\label{rootordering}
|u_{\rm h}^{(1,2)}| > |u_k| > u_{\rm h}^{(j)} \quad (j\neq1,2)\, .
\end{equation}
%


\subsection{The Counting Function and the NLIE}
\label{sec:counting}

A nice way to exploit the existence of the hidden hole degrees
of freedom employs the so-called
{\it counting function}, see \cite{Feverati:2006tg} and references
therein. It is defined as
\begin{equation}\label{countingf}
Z(u)=L\,\phi(u,\half)+\sum_{k=1}^M\phi(u-u_k,1)\quad \mbox{where}\quad
\phi(u,\xi)=i\,\log\left(\frac{i\xi+u}{i\xi-u}\right).
\end{equation}
Its name stems from the fact that, as one immediately sees
from the definition \eqref{countingf},
$Z(\pm\infty)=\pm \,\pi\,(L+M)$ while
the Bethe equations for the magnons and holes may
be, respectively, expressed as
\begin{eqnarray}\label{cfbeqD}
Z(u_j)&=&\pi\,(2\,n_j+\delta-1) \qquad j=1,\ldots, M\, ,  \\
\label{cfbeqZ}
Z(u_{\rm h}^{(k)})&=&\pi\,(2\,n_{\rm h}^{(k)}+\delta-1) \qquad k=1,\ldots, L\, ,
\end{eqnarray}
where
\begin{equation}
\delta=L+M \quad\mbox{mod}\quad 2\, .
\end{equation}
So $Z(u)$ is a smooth function which yields the corresponding
mode number (times $\pi$) whenever $u$ equals a
hole or magnon root. The mode numbers clearly ``label''
or  ``count'' the solutions of the Bethe equations, and
the counting function smoothly interpolates between them.

To write down the one-loop non-linear integral equation, we recall
\cite{Feverati:2006tg} that for an arbitrary function $f(u)$,
which is analytic within a strip around the real axis, the following identity holds
\begin{equation} \label{cruciali}
\sum_{k=1}^M
f(u_k)+\sum_{j=1}^L\,f(u_{\rm h}^{(j)})=-\int_{-\infty}^\infty\frac{du}{2\pi}\,f'(u)\,Z(u)+\int_{-\infty}^\infty
\frac{du}{\pi}\,f'(u)\,\mbox{Im}\log\left[1+(-1)^\delta\,e^{i\,Z(u+i\,0)}\right].
\end{equation}
Applying this identity to $Z(u)$ and adapting the
steps of \cite{Feverati:2006tg} to the present case,
we find\footnote{Due to superficial divergencies one needs to apply \eqref{cruciali} to $Z''(u)$ and then to integrate twice the resulting equation. The constants of integration are fixed by antisymmetry of $Z(u)$ and the condition
\begin{equation}\label{cond1}
\lim_{u \to \infty} Z'(u)=0\, .
\end{equation}
}
\begin{eqnarray}\label{Zuspace}
\nonumber Z(u)&=&i\,L\,\log\frac{\Gamma\left(1/2+i\,u\right)}{\Gamma\left(1/2-i\,u\right)}+\sum_{j=1}^L\,i\,\log\frac{\Gamma\left(-i\,(u-u_{\rm h}^{(j)})\right)}{\Gamma\left(i\,(u-u_{\rm h}^{(j)})\right)}\\
&+&\lim_{\alpha \to \infty}\int_{-\alpha}^\alpha \frac{dv}{\pi}\,i\,\frac{d}{du}\log\frac{\Gamma(-i\,(u-v))}{\Gamma(i\,(u-v))}\,\mbox{Im}\log\left[1+(-1)^\delta\, e^{i\,Z(v+i\,0)}\right].
\end{eqnarray}

The identity \eqref{cruciali} may also be used to express all conserved charges in terms of the counting function. The first charge (the momentum), however, needs to be regularized
\begin{equation}
P=\lim_{\alpha \to \infty}\left(-\int_{-\alpha}^\alpha \frac{du}{2\pi}\,p'(u)\,Z(u)-\sum_{j=1}^L
p(u_{\rm h}^{(j)})+\int_{-\alpha}^\alpha \frac{du}{\pi}\,p'(u)\,\mbox{Im}\log\left[1+(-1)^\delta\,e^{i\,Z(u+i\,0)}\right]\right).
\end{equation}
In the above formula $p(u)$ denotes the momentum of a single particle
\begin{equation}
p(u)=\frac{1}{i}\,\log\frac{u+i/2}{u-i/2}.
\end{equation}
Due to antisymmetry of $Z(u)$ and $p(u)$ one easily finds
\begin{equation}
P=0.
\end{equation}
Similarly, the one-loop anomalous dimension $\gamma_1$,
see \eqref{1loopmomeng}, may be rewritten as
\begin{eqnarray}\label{energy}
\nonumber
\gamma_1&=&
4\,\gamma_{\rm E}\, L+2\,\sum_{j=1}^L\left\{\psi(1/2+i\,u^{(j)}_{\rm h})+\psi(1/2-i\,u^{(j)}_{\rm h})\right\}\\
&+&2\,\int_{-\infty}^\infty\frac{dv}{\pi}\,i\,\frac{d^2}{dv^2}\,\left(\log\frac{\Gamma\left(1/2+i\,v\right)}{\Gamma\left(1/2-i\,v\right)}\right)\mbox{Im}\log\left[1+(-1)^\delta\,e^{i\,Z(v+i\,0)}\right],
\end{eqnarray}
where $\gamma_{\rm E}$ is Euler's constant.

Note that the NLIE \eqref{Zuspace} in conjunction with the
Bethe equations for the hole roots \eqref{cfbeqZ} is
fully equivalent, for the ground-state, to the algebraic
Bethe equations \eqref{sl2} for arbitrary finite values
of $M$ and $L$. (The generalization to
the case of excited states is fairly straightforward but
will not be discussed in this paper.) Likewise,
the expressions for the one-loop
anomalous dimension $\gamma_1$
given in \eqref{1loopmomeng} and \eqref{energy}
are equivalent.


\subsection{Magnon Density}
\label{sec:magnondensity}

If the number of magnon roots $M$ gets large we may expect,
for the groundstate, that they form a dense distribution on the
union of two intervals $[-b,-a]$ and $[a,b]$ on the
real axis. This allows us to introduce a distribution density $\magnonrho(u)$,
see section 3.2.~of \cite{Eden:2006rx} for further details.
It then follows from \eqref{modenumbers} and \eqref{cfbeqD} that
\begin{equation} \label{reltodens}
\frac{1}{M}\,\frac{d}{du} Z(u)= 2\,\pi \,\magnonrho(u)+2\,\pi\,\frac{L-2}{M}\,\delta(u)+\Op\left(\frac{1}{M^2}\right),
\quad {\rm with} \quad
2 \int_a^b du\,\magnonrho(u)=1\, ,
\end{equation}
where the $\delta$-function stems from the gap in the
center of the magnon mode numbers \eqref{modenumbers}.
Using this relation one can rewrite \eqref{countingf} as
\begin{equation}\label{largeMeq}
2\,\pi\,\magnonrho(u)+2\,\pi\,\frac{L-2}{M}
\delta(u)
-\frac{L}{M} \frac{1}{u^2+\frac{1}{4}}
-2
\left(\int_{-b}^{-a} dv + \int_a^b dv\right)
\frac{\magnonrho(v)}{(u-v)^2+1}
=0\, ,
\end{equation}
where $u\in [-b,-a] \cup [a,b]$. If there is a gap $2a>0$
we therefore may drop the term involving the $\delta$-function in
\eqref{largeMeq}.
In principle, if interpreted appropriately, this equation should hold
for large $M$ and {\it arbitrary}, small or large, $L$.

If in addition $L$ stays finite (but arbitrary) we can
apply the scaling procedure $\bar u=u/M$ of \cite{Eden:2006rx}, as in
this case the gap $2 a$ closes ($a \rightarrow 0$), and
in addition $b \rightarrow M/2$.
Then the non-singular integral equation \eqref{largeMeq} turns
into the singular integral equation, and with
$\korbarrho(\bar{u}) =M \,\magnonrho(u)$ we find
\begin{equation}\label{1loopsingular2}
-4\,\pi\,\delta(\bar u)
-2\,\pint_{-\half}^{\half} d\bar u'\,
\frac{\korbarrho(\bar u')}{(\bar u-\bar u')^2}
=0\, .
\end{equation}
The solution is the (singular) density
\begin{equation}\label{1loopdenssol}
\korbarrho(\bar u)=
\frac{1}{\pi}\,\log \frac{1+\sqrt{1-4\,\bar u^2}}{1-\sqrt{1-4\,\bar u^2}}=
\frac{2}{\pi}\,\text{arctanh}\left(\sqrt{1-4\,\bar u^2}\right)\, ,
\end{equation}
first derived in \cite{Korchemsky:1995be}. It should be
considered as a distribution (in the mathematical sense)
rather than as a regular function. The reason is that
the expression for the one-loop anomalous dimension
in \eqref{1loopmomeng} formally turns into
$4 \pi \int d\bar u\, \magnonbarrho(\bar u) \delta(\bar u)=4\pi \magnonrho(0)=\infty$.
However, a more careful analysis \cite{Eden:2006rx} of the
multiplication of the distributions
$\magnonrho(\bar u)$ and $\delta(\bar u)$ leads to
\begin{equation}\label{1loopscaling}
\gamma_1=8\,\log M+\Op(M^0)\, .
\end{equation}

If instead $L \rightarrow \infty$ along with $M \rightarrow \infty$
such that $\beta=M/L$ is kept finite, the gap $2a$ does not close.
We may then drop the $\delta$-function term in \eqref{largeMeq}
and obtain after rescaling $\bar u=u/M$ with
$\bar a=a/M$, $\bar b=b/M$
\begin{equation}\label{precisioneq}
-\frac{1}{\beta} \frac{1}{{\bar u}^2}
-2\left(\pint_{-\bar b}^{-\bar a} d\bar v + \pint_{\bar a}^{\bar b}
d\bar v\right)
\frac{1}{(\bar u-\bar v)^2}
\, \magnonbarrho(\bar v)=0\, ,
\end{equation}
which is essentially (up to a rescaling of $\bar u$ by $\beta$)
the derivative $d/d \bar u$ of the singular two-cut
integral equation first derived in \cite{Beisert:2003ea}.
The original equation is easily reconstructed by integrating both
sides of \eqref{precisioneq} w.r.t.~$\bar u$, with a constant
of integration of $2 \pi/\beta$ on the right hand side.
The explicit solution for the density $\magnonbarrho(\bar v)$
along with $\bar a,\bar b$ was also given in \cite{Beisert:2003ea}.
When $\beta \rightarrow \infty$ the gap $2\bar a$ disappears
and the limiting distribution \eqref{1loopdenssol} is recovered.
However, this procedure does {\it not} reproduce the correct
behavior of the anomalous dimension of the previous
large $M$ limit at fixed $L$,
i.e.~\eqref{1loopscaling}; instead, one finds
\begin{equation}\label{1loopspinning}
\gamma_1=\frac{8}{L}\,\log^2 \frac{M}{L}+\ldots\, .
\end{equation}
See also the discussion in \cite{Eden:2006rx}.
We notice that large $M$ analysis is quite subtle if the
gap $2 a$ is very small but non-vanishing.

In fact, there is a very interesting perturbation on the
scaling behavior \eqref{1loopscaling} first noticed in
\cite{Belitsky:2006en}.
Let us understand this effect by a  more refined analysis
of \eqref{reltodens}, \eqref{largeMeq}.
It is convenient to split the density $\magnonrho(u)$ into
the singular, leading piece $\korrho(u)$ and a
fluctuation correction $\tilde \sigma(u)$:
$\magnonrho(u)=\korrho(u)+\tilde \sigma(u)$ where
$\korrho(u)=1/M\, \korbarrho(u/M)$, see \eqref{1loopdenssol}.
The trick is to now add
\begin{equation}\label{trick}
2 \int_{-a}^{a} dv \frac{\korrho(v)}{(u-v)^2+1}
=\frac{4\,\log M}{\pi\, M}\left(\arctan(u+a)-\arctan(u-a)\right)+ \Op(M^0)
\end{equation}
to \eqref{largeMeq}. We then see that
$\tilde \sigma(u)$ scales as $\log M/M$ and we should therefore define,
in analogy with \cite{Eden:2006rx}, a {\it fluctuation density}
$\sigma(u)$ through
\begin{equation}\label{fluct}
\magnonrho (u)=\korrho (u)-\frac{8\,\log M}{M}\,\sigma(u)\, .
\end{equation}
It satisfies
\begin{eqnarray}\label{flucteq}
&&2\,\pi\,\sigma(u)
-\frac{1}{2\pi} \left(\arctan(u+a)-\arctan(u-a)\right)
+\frac{\LoverlogM}{8}\,\frac{1}{u^2+\frac{1}{4}}\\\nonumber
&&-2
\left(\int_{-\infty}^{-a} dv + \int_a^{\infty} dv\right)
\frac{\sigma(v)}{(u-v)^2+1}=0\, .
\end{eqnarray}
This integral equation fully determines the fluctuation
density $\sigma(u)$, as the edge parameter $a$ may
be determined from the normalization condition $\left(\int^{-a}_{-\infty}+\int^{\infty}_{a} \right)du \rho(u)=1$, which implies
\begin{equation}\label{normalization}
\LoverlogM=\frac{4\,a}{\pi}-8\,\int^{a}_{-a} du \,\sigma(u)
\end{equation}
The one-loop anomalous
dimension is then given from \eqref{1loopmomeng} by
\begin{equation}\label{flucteng}
\frac{\gamma_1(\LoverlogM)}{\log M}=
8-\frac{16}{\pi}\,\arctan{2a}-16\,\left(\int_{-\infty}^{-a}\,du + \int_a^{\infty}\,du\right)
\frac{\sigma(u)}{u^2+\frac{1}{4}}\, .
\end{equation}
%

\subsection{Fourier Space Equation}
\label{sec:fourier}

It is very instructive to change from $u$-space to
Fourier space. After rewriting \eqref{flucteq} as
\begin{eqnarray}\label{rewritting}
\sigma(u)&=&\frac{1}{4\pi^2}\,\left(\arctan(u+a)-\arctan(u-a)\right)-\frac{\LoverlogM}{16\,\pi}\frac{1}{u^2+\frac{1}{4}} \\ \nonumber
&+& \int^{\infty}_{-\infty} \frac{dv}{\pi}\frac{\sigma(v)}{1+(u-v)^2}-\int^{a}_{-a} \frac{dv}{\pi} \frac{\sigma(v)}{1+(u-v)^2}
\end{eqnarray}
and Fourier transforming\footnote{%
We have included a factor of $e^{-\frac{t}{2}}$ into this
definition for convenience. For all other Fourier transformed quantities
in this paper, in particular all kernels $\hat K$, we do not
include such a factor.
}%
\begin{equation}\label{Fourier}
\hat{\sigma}(t)=
e^{-\frac{t}{2}} \int_{-\infty}^{\infty} du\,e^{-i\,t\,u}\,\sigma(u)
\end{equation}
one obtains
\begin{equation}\label{fouriermagnons}
\hat{\sigma}(t)=\frac{t}{e^t-1}\left(\hat{K}_{\rm h}(t,0;a)-\frac{\LoverlogM}{8\,t}\,-4\,\int^{\infty}_0 dt' \,\hat{K}_{\rm h}(t,t';a)\,\hat{\sigma}(t')\right),
\end{equation}
where the kernel is given by
\begin{equation}\label{ktildeIR}
\hat{K}_{\rm h}(t,t';a)
=\frac{e^{\frac{t'-t}{2}}}{4\,\pi\,t}\,\int^{a}_{-a}du\,\cos(t\,u)\,\cos(t'\,u),
\end{equation}
which leads to the expression \eqref{ktilde} in the introduction.
Likewise, Fourier-transforming the normalization condition
\eqref{normalization} yields the relation \eqref{normalizationtspace}
between the physical parameter $\LoverlogM$ and
the fluctuation density $\hat \sigma(t)$ in Fourier space
stated already in the introduction.
The one-loop anomalous dimension is then given by
\begin{eqnarray}
\frac{\gamma_1(\LoverlogM)}{\log M}&=&8\,\left[1-\frac{2}{\pi}\,\arctan{2a} \right.\\
\nonumber &&\quad \ -\left. 4\,\int^{\infty}_{0}dt \,\left(\hat{\sigma}(t)-4\,t\,\int^{\infty}_{0} dt'\, \hat{K}_{\rm h}(t,t';a)\,\hat{\sigma}(t') \right) \right].
\end{eqnarray}
%

\subsection{Hole Density}
\label{sec:holedensity}

The Bethe roots corresponding to the small holes lie inside
some interval $[-\holegap,\holegap]$. In the ``thermodynamic'' limit
$L \rightarrow \infty$, where the number of small holes tends
to infinity, their one-loop root distribution density $\holerho(u)$ is
related to the counting function through
\begin{equation}\label{rhohole}
\frac{1}{L}\,\frac{d}{du}\,Z(u)=2\,\pi\,\holerho(u)
+\Op\left(\frac{1}{L}\right),
\quad {\rm with} \quad
\int_{-c}^c du\,\holerho(u)=1\, ,
\end{equation}
as one easily derives from \eqref{modenumbersholes}
and \eqref{cfbeqZ}.
Using \eqref{Zuspace}, we may then derive a nonlinear integral equation for
the distribution of holes
\begin{eqnarray}
\holerho(u)&=&
\frac{1}{L}\,
\big(\psi(i\,(u-u_{\rm h}^{(1)}))+\psi(-i\,(u-u_{\rm h}^{(1)})
+\psi(i\,(u+u_{\rm h}^{(1)}))+\psi(-i\,(u+u_{\rm h}^{(1)}))\big)
\nonumber \\
& &
+\frac{1}{L}\,\frac{d}{du}{\cal I}(u)
-\frac{1}{2\,\pi}\,\big(\psi(\half+i\,u)+\psi(\half-i\,u)\big)
\nonumber \\
& &
+\int_{-\holegap}^{\holegap}\frac{dv}{2\pi}\,\big(\psi(i\,(u-v))+\psi(-i\,(u-v))\big)\,
\holerho(v)\, ,
\label{1looprho}
\end{eqnarray}
where the term $\frac{1}{L}\,\frac{d}{du}{\cal I}(u)$ denotes
the derivative of the last line in \eqref{Zuspace}.
The terms on the r.h.s.~of the first line of \eqref{1looprho} are the
contributions of the two large holes with rapidities
$u_{\rm h}^{(1)},u_{\rm h}^{(2)}=-u_{\rm h}^{(1)}$,
{\it cf} \eqref{rootordering},
where we have also implicitly assumed $L \ll M$. Then the
two rapid holes behave as\footnote{
Extensive numerical studies indicate that for the ground state
at large $M$ all charges $q_i$ in \eqref{trm} are small
except $q_2$. Then one finds from $t(u)=0$ and
\eqref{q2} $u_{\rm h}^{(1,2)}\simeq \pm M/\sqrt{2}$.
See also \cite{Belitsky:2006en}.}
$u_{\rm h}^{(1,2)}\simeq \pm M/\sqrt{2}$, while the 
term $1/L\,\frac{d}{du}{\cal I}(u)$ in \eqref{1looprho} yields merely an additive $2\,\log 2$, see appendix \ref{app:nonlinearterm} for a 
discussion of this point. The four terms 
on the r.h.s.~of the first line of \eqref{1looprho} thus
behave like $4 \log M/\sqrt{2}$. Using \eqref{newlimit}
we thus derive a {\it linear} integral equation
\begin{equation}\label{1looprhoscaled}
\holerho(u)=
\frac{2}{\pi \LoverlogM}
-\frac{1}{2\pi}\,\big(\psi(\half+i\,u)+\psi(\half-i\,u)\big)
+\int_{-\holegap}^{\holegap}\frac{dv}{2\pi}\,\big(\psi(i\,(u-v))+\psi(-i\,(u-v))\big)
\holerho(v)\, .
\end{equation}
One then finds the generalized one-loop scaling function,
c.f.~\eqref{newlog}, from \eqref{energy}
\begin{equation}\label{1loopandim}
\frac{\gamma_1(\LoverlogM)}{\log M}=
8+2\,  \LoverlogM\,\int_{-\holegap}^{\holegap}du\,\holerho(u)\,
\big(\psi(\half+i\,u)+\psi(\half-i\,u)-2\,\psi(1)\big).
\end{equation}

In order to easily generate the series expansion of
\eqref{1loopandim} in powers of  $\LoverlogM$, defined in
\eqref{newlimit},
it is useful to rescale $u$ and define
\begin{equation}\label{rescale}
\bar u = \frac{u}{\holegap}
\qquad {\rm and} \qquad
\holebarrho(\bar u)=  \LoverlogM\,\holegap\, \holerho(u).
\end{equation}
Defining the non-singular kernel
\begin{equation}\label{1loopholekernel}
K(\bar u,\bar v)=\frac{\holegap}{2\pi}\,
\bigg(
\psi(i\,\holegap\,(\bar u-\bar v))+\psi(-i\,\holegap\,(\bar u-\bar v))
-\psi(\half+i\,\holegap\,\bar u)-\psi(\half-i\,\holegap\,\bar u)
\bigg)\, ,
\end{equation}
the integral equation \eqref{1looprhoscaled} becomes
\begin{equation}\label{1loopbarrho}
\holebarrho(\bar u)=
\frac{2}{\pi}\,\holegap
+\int_{-1}^1 d \bar v\, K(\bar u,\bar v)\,
\holebarrho(\bar v)\, .
\end{equation}
It is of Fredholm-type and may be immediately expanded in
the small parameter $\holegap$ and iteratively solved as a power
series in $\holegap$. The relation to the parameter  $\LoverlogM$ is then
determined through the normalization condition in
\eqref{rhohole} which becomes
\begin{equation}\label{1loopbarrhonorm}
 \LoverlogM=\int_{-1}^1dv\, \holebarrho(\bar u)\, .
\end{equation}
This yields  $\LoverlogM$ as a series in $\holegap$.
The generalized one-loop scaling function \eqref{1loopandim}
becomes
\begin{equation}\label{1loopbarandim}
\frac{\gamma_1(\LoverlogM)}{\log M}=
8+2\,\int_{-1}^1d \bar u\,\holebarrho(\bar u)\,
\big(\psi(\half+i \holegap \bar u)+\psi(\half-i \holegap \bar u)-2\,\psi(1)\big).
\end{equation}
This yields the one-loop scaling function as a series in $\holegap$.
Inverting the series \eqref{1loopbarrhonorm} and substituting
into the expansion of \eqref{1loopbarandim} gives
the desired series of the scaling function in terms of  $\LoverlogM$.
It starts out as
\begin{eqnarray}
\frac{\gamma_1(\LoverlogM)}{\log M}&=&
8-8\,\LoverlogM\,\log 2+\frac{7}{12}\,\LoverlogM^3\,\pi^2 \,\zeta(3)
-\frac{7}{6}\,\LoverlogM^4 \,
  \pi ^2 \, \log 2\, \zeta(3)
\nonumber \\ \label{ellexpansion}
&+&
2 \,\LoverlogM^5 \, \left(\frac{7}{8} \, \pi ^2 \, \log^2 2\, \zeta(3)-\frac{31}{640} \, \pi ^4\,
  \zeta(5)\right)+
  \Op(\LoverlogM^6).
\end{eqnarray}
Note that by analytic continuation the density of the holes is related to $\sigma(u)$ via
\begin{equation}\label{holessigma}
\LoverlogM \,\holerho (u)=\frac{2}{\pi}-8\,\sigma(u) \qquad \qquad u\in (-c,c)
\end{equation}
which may be rewritten as
\begin{equation}\label{holessigmatspace}
\LoverlogM \,\holerho (u)=\frac{2}{\pi}-\frac{8}{\pi}\,\int^{\infty}_{0}dt\, \hat{\sigma}(t)\,e^{\frac{t}{2}} \,\cos{tu}\, .
\end{equation}

The preceding derivation proceeds from the counting function,
{\it cf} \eqref{rhohole}. We will closely follow this procedure
in the next chapter \ref{sec:all-loop}, where we will treat
the higher-loop case. It should be noted, however, that
our solution 
\eqref{1loopbarrho},\eqref{1loopbarrhonorm},\eqref{1loopbarandim}
may also be immediately recovered by Fourier analyzing
the results of the previous section \ref{sec:fourier}.
The reader should multiply \eqref{fouriermagnons} with 
$e^{\frac{t}{2}} \,\cos{tu}$, integrate in $t$ over the positive real semi-axis and use the integral representation of the kernel \eqref{ktildeIR}. Subsequently rewriting $\LoverlogM$ in terms of 
$\oldtildesigma(t)$ with 
the help of \eqref{normalizationtspace} and finally using the relation \eqref{holessigmatspace} it is straightforward to derive \eqref{1looprhoscaled}. We thus conclude that
\begin{equation}
a=c\, .
\end{equation}
This equation tells us that the gap $[-a,a]$ in the distribution 
of magnon roots is densely filled by the (small) hole roots.

\section{All-Loop Theory}
\label{sec:all-loop}
\subsection{The Asymptotic Non-Linear Integral Equation (NLIE)}
\label{sec:asympnlie}

Let us now extend the one-loop results of the last chapter
to the higher loop case. We will use the asymptotic Bethe
ansatz for AdS/CFT, based on the S-matrix approach
\cite{Staudacher:2004tk}.
In the $\mathfrak{sl}(2)$ subsector the asymptotic all-loop Bethe equations \cite{Beisert:2005fw,Beisert:2006ez} read
\begin{equation}\label{sl2eq}
\left(\frac{x_k^+}{x_k^-}\right)^L=\prod_{j\neq
k}^M\frac{u_k-u_j-i}{u_k-u_j+i}\left(\frac{1-\frac{g^2}{x_k^+x_j^-}}{1-\frac{g^2}{x_k^-x_j^+}}\right)^2 e^{2\,i\,\theta(u_k,u_j)}.
\end{equation}
We define the all-loop asymptotic counting function as
\begin{eqnarray}\label{allloopZ}
\nonumber Z(u)&=&i\,L\,\log\frac{x(i/2+u)}{x(i/2-u)}+i\,\sum^{M}_{k=1}
\log\frac{i+u-u_k}{i-(u-u_k)}\\
&-&2\,i\,\sum^{M}_{k=1}
\log\frac{1+\frac{g^2}{x(i/2+u)x(i/2-u_k)}}{1+\frac{g^2}{x(i/2-u)x(i/2+u_k)}}+\sum^{M}_{k=1}
\theta (u,u_k).
\end{eqnarray}
As in the one-loop case, one finds the corresponding 
non-linear integral equation
\begin{eqnarray}\label{allloopZv2}
\nonumber Z(u)&=&i\,L\,\log\frac{x(i/2+u)}{x(i/2-u)}+\int_{-\infty}^\infty\frac{dv}{2\pi}\,\phi'(u-v,1)\,Z(v)\\
\nonumber&-&\sum_{j=1}^L\,\phi(u-u_{\rm h}^{(j)},1)-\int_{-\infty}^\infty\frac{dv}{\pi}\,\phi'(u-v,1)\,\mbox{Im}\log\left[1+(-1)^\delta \,e^{i\,Z(v+i\,0)}\right]\\
\nonumber&+&\int_{-\infty}^\infty\frac{dv}{2\pi}\,\left(2\,i\,\frac{d}{dv}\log\frac{1+\frac{g^2}{x(i/2+u)x(i/2-v)}}{1+\frac{g^2}{x(i/2-u)x_(i/2+v)}}-\theta(u,v)\right)\,Z(v) \\ \nonumber
&+&\sum_{j=1}^L \left(2\,i\,\log\frac{1+\frac{g^2}{x(i/2+u)x(i/2-u_{\rm h}^{(j)})}}{1+\frac{g^2}{x(i/2-u)x_(i/2+u_{\rm h}^{(j)})}}-\theta(u,u_{\rm h}^{(j)})\right)\\
&-&\nonumber\int_{-\infty}^\infty\frac{dv}{\pi}\,\left(2\,i\,\frac{d}{dv}\,\log\frac{1+\frac{g^2}{x(i/2+u)x(i/2-v)}}{1+\frac{g^2}{x(i/2-u)x_(i/2+v)}}-\theta(u,v)\right)\,\mbox{Im}\log\left[1+(-1)^\delta \,e^{i\,Z(v+i\,0)}\right].
\\
\end{eqnarray}
The counting function defined in \eqref{allloopZ} satisfies a similar relation to \eqref{reltodens}, but with the all-loop density on the 
r.h.s. 

\subsection{The NLIE in Fourier Space}
\label{sec:NLBES}

In Fourier $t$-space equation \eqref{allloopZv2} becomes
\begin{eqnarray}\label{Ztspace}
\hat{Z}(t)&=&\frac{2 \,\pi \,L
e^{\frac{t}{2}}}{i\,t\,(e^{t}-1)} J_{0}(2gt)-\sum^{L}_{j=1}
\frac{2\,\pi\,\cos \left(t\,u^{(j)}_{\rm h} \right)}{i\,t\,(e^t-1)} -\frac{2}{e^t-1} \hat{\cal L}(t)\nonumber \\
&+&8\,g^2 \frac{e^{\frac{t}{2}}}{e^t-1} \int^{\infty}_{0}\,dt'\,
e^{-\frac{t'}{2}}\,\hat{K}(2gt,2gt')\,\bigg(t'\,\hat{\cal L}(t')
\nonumber\\
&+&\frac{\pi}{i}
\sum^{L}_{j=1} \cos\left(t'\,u^{(j)}_{\rm h}\right) \bigg) \nonumber \\
&-&4
\,g^2\,\frac{e^{\frac{t}{2}}}{e^{t}-1}\int^{\infty}_{0}\,dt'\,e^{-\frac{t'}{2}}\,t'\,
\hat{K}(2gt,2gt')\,\hat{Z}(t'),
\end{eqnarray}
where $\hat{\cal L}(t)$ denotes the Fourier transform of the ``Im log'' term. Note that the $\hat{Z}(t)$ has a first order pole at $t=0$. This in accordance with \eqref{allloopZ}, since the Fourier transform of this expression must be understood in the principal value sense.
Note that we have not made any approximations. Therefore
\eqref{Ztspace} is still fully equivalent to the orginal set
of discrete asymptotic equations \eqref{sl2eq}.

\subsection{Large Parameter Integrals}
\label{sec:aexp}

Let us now investigate the effects of taking the large $M$ limit
with $L \ll M$.
It will be important to understand the large $M$ expansion of  
integrals of the form
\begin{equation}\label{fintegral}
f(M)=\int^{\infty}_{0}dx\, h(x)\,\sin \left(u(M)\, x \right),
\end{equation}
where $h(x)$ is a smooth integrable function on 
$[ 0,\infty)$
and $u(M)\to \infty$ when $M\to \infty$.
We first note that because of the relation to the
Fourier transform (Plancherel's theorem) $\lim_{M \to \infty} f(M)=0$.
Since $f(M)$ is meromorphic and vanishes at infinity we have
\begin{equation}\label{fatlargeM}
f(M)=\sum^{\infty}_{j=0} \frac{c_j}{u(M)^{1+j}}\, .
\end{equation}
To find $c_0$ it is sufficient to note that
\begin{equation}\label{c0coeff}
c_0=\lim_{M \to \infty} u(M) \, f(M)=\lim_{M \to \infty}
\int^{\infty}_{0} dx \,h(x)\left(-\frac{d}{dx} \cos \left(u(M)\,
x\right)\right)=h(0),
\end{equation}
since the integral after a partial integration vanishes again. By subsequent integrations by part one finds that
\begin{equation}\label{cncoeff}
c_n=\lim_{M \to \infty} u(M)^{n+1} \, \left(f(M)-\sum^{n-1}_{j=1}
\frac{c_j}{u(M)^{1+j}}\right)=(-1)^{\frac{n}{2}} \, h^{(n)}(0)
\qquad \mbox{for even $n$}\, .
\end{equation}
The odd $c_n$ coefficients vanish, as follows from \eqref{fintegral}.

\subsection{The Leading Order Equation}
\label{sec:lo}

To derive from \eqref{Ztspace} an equation reproducing the leading contribution to the scaling function in the limit where $M\to \infty$ and $L$ is kept fixed, it is sufficient to observe, based on the results of the previous subsection, that upon iterating \eqref{Ztspace} only terms of the form
\begin{equation}\label{theterm}
\frac{2 \,\pi \,
e^{\frac{t}{2}}}{i\,t\,(e^{t}-1)}-
\frac{2\,\pi\,\cos \left(t\,u^{(1,2)}_{\rm h} \right)}{i\,t\,
(e^t-1)}\,,
\end{equation}
where $u^{(1,2)}_{\rm h} \simeq \pm \sqrt{\frac{1}{2}\,q_2} \simeq \pm \frac{M}{\sqrt{2}}$ represent the universal holes, will give the leading (logarithmic) contribution. This is because we have
\begin{equation}\label{smallholesvanish}
u_{\rm h}^{(j)}\simeq 0 \qquad j=3,\ldots, L
\end{equation}
at leading order, and the terms involving $\hat{\cal L}(t)$ do not contribute at this order, see appendix \ref{app:nonlinearterm}.
Thus the leading all-loop equation reads
\begin{eqnarray}\label{alleq}
\nonumber \hat{Z}(t)&=&\frac{4 \,\pi \,
e^{\frac{t}{2}}}{i\,t\,(e^{t}-1)}-
\frac{4\,\pi\,\cos \left(t\,u^{(1)}_{\rm h} \right)}{i\,t\,(e^t-1)}\\
&-&4
\,g^2\,\frac{e^{\frac{t}{2}}}{e^{t}-1}\,\int^{\infty}_{0}\,dt'\,e^{-\frac{t'}{2}}\,t'\,
\hat{K}(2gt,2gt')\,\hat{Z}(t')
\end{eqnarray}
Upon subtracting the one-loop part of this equation
\begin{equation}\label{split}
\hat{Z}(t)=\hat{Z}_0(t)+\delta \hat{Z}_{\mbox{\tiny BES}}(t)
\end{equation}
and identifying $\delta \hat{Z}(t)$ with the fluctuation density
\begin{equation}\label{deltaZdensity}
\delta\hat{Z}_{\mbox{\tiny BES}}(t)=16\,\pi\,i\,g^2\,e^{\frac{t}{2}}\,
\frac{\oldtildesigma_{\mbox{\tiny BES}}(t)}{t}\, \log(M)
\end{equation}
one rederives the equation of \cite{Beisert:2006ez}:
\begin{equation}\label{esbes}
\oldtildesigma_{\mbox{\tiny BES}}(t)=\frac{t}{e^t-1}\,\left(\hat{K}(2gt, 0)-4\,g^2\,\int^{\infty}_{0}\,dt'\,\hat{K}(2gt,2gt')\,
\oldtildesigma_{\mbox{\tiny BES}}(t')\right).
\end{equation}
%

\subsection{Subleading Corrections to the Twist Operator Dimensions}
\label{sec:subleading}

The large $M$ expansion of the anomalous dimensions of twist operators is expected to have the following form
\begin{equation}\label{gammaexp}
\gamma=f(g)\,\log{M}+f_{{\rm sl}}(g,L)+\mathcal{O}\left(\frac{1}{(\log{M})^2}\right),
\end{equation}
where $f_{{\rm sl}}(g,L)$ denotes the subleading effects
of $\Op(M^0)$.
These are easily obtained from \eqref{Ztspace}, and we may compute $f_{{\rm sl}}(g,L)$ to arbitrary order of perturbation theory:
\begin{eqnarray}\label{subleading}
\nonumber f_{{\rm sl}}(g,L)&=&\left(\gamma-(L-2)\,\log 2\right)\,f(g)-8\,(7-2\,L)\,\zeta(3)\,g^4\\\nonumber
&+&8\,\left(\frac{4-L}{3}\,\pi^2\,\zeta(3)+(62-21\,L)\,\zeta(5)\right)\,g^6\\
\nonumber&-&\frac{8}{15}\,\left((13-3\,L)\,\pi^4\,\zeta(3)+5\,(32-11\,L)\,\pi^2\,\zeta(5)+75\,(127-46\,L)\,\zeta(7)\right)\,g^8\\
&\pm&\ldots
\end{eqnarray}
Notice that the ``universality'', i.e.~$L$-independence of the scaling function $f(g)$ is lost when one computes these $\Op(M^0)$
terms. They contain $L$-independent and terms linear in $L$. 

\section{The Generalized Scaling Function}
\label{sec:genscaling}

\subsection{Derivation}

Let us now finally treat the novel scaling limit \eqref{newlimit}, i.e.~we
consider the limit $L,M\to\infty$ with $\LoverlogM=L/\log M$ kept fixed. In this limit, in contradistinction to section \ref{sec:lo}, also the $L-2$ remaining holes contribute. Although individual hole terms separately do not develop logarithmic behavior in $M$, their collective
contribution will be proportional to $L= \LoverlogM \log M$. 
Furthermore, in this limit all terms involving $\hat{\cal L}(t)$ can be dropped, see appendix \ref{app:nonlinearterm}.  Thus 
\eqref{Ztspace} for the counting function $\hat{Z}(t)$ in Fourier space
linearizes in this limit to the form
\begin{eqnarray}\label{leadingFTlimit}
\hat{Z}(t)&=&\frac{2 \,\pi \,L
e^{\frac{t}{2}}}{i\,t\,(e^{t}-1)}J_0(2gt)-\sum^{L}_{j=1}
\frac{2\,\pi\,\cos \left(t\,u^{(j)}_{\rm h} \right)}{i\,t\,(e^t-1)}\nonumber \\
&+&8\pi\,\,g^2\, \frac{e^{\frac{t}{2}}}{i\,(e^t-1)} \sum^{L-2}_{j=1}\,\int^{\infty}_{0}\,dt'\,
e^{-\frac{t'}{2}}\,\hat{K}(2gt,2gt') \cos\left(t'\,u^{(j)}_{\rm h}\right) \nonumber \\
&-&4
\,g^2\,\frac{e^{\frac{t}{2}}}{e^{t}-1}\,\int^{\infty}_{0}\,dt'\,e^{-\frac{t'}{2}}\,t'\,
\hat{K}(2gt,2gt')\,\hat{Z}(t').
\end{eqnarray}
Note that in above formula only quantum corrections to $u_{\rm h}^{j}$ for $j=3,\ldots,L$ need to be taken into account, since the corrections to the universal holes are, upon the iteration, subleading. 
In similarity to section \ref{sec:lo} we strip off the one-loop part 
by defining
\begin{equation}\label{stripoffFT}
\hat{Z}(t)=\hat{Z}_0(t)+\delta \hat{Z}(t).
\end{equation}
We relate $\delta \hat{Z}(t)$ to the fluctuation density 
$\oldtildesigma(t)$ through
\begin{equation}\label{split2}
\delta \hat{Z}(t)=16\,\pi\, i\,e^{\frac{t}{2}}\,\frac{\oldtildesigma(t)}{t} \, \log {M}\, ,
\end{equation}
and derive to the desired order
\begin{eqnarray}\label{densityAM}
\nonumber\oldtildesigma(t)=\frac{t}{e^t-1}\,&& \left[g^2\,\hat{K}(2gt,0)-\frac{\LoverlogM}{8}\frac{J_0(2gt)}{t}+\frac{1}{8\log M}\,\sum_{j=3}^{L}\frac{e^{-t/2}\cos(t\,u_{\rm h}^{(j)})}{t}\right.\\
&&\ -\left.
\frac{g^2}{2}\,\frac{1}{\log M}\,\sum_{j=3}^{L}\,\int_0^\infty\,dt'\,\hat{K}(2gt,2gt')\,e^{-t'/2}\,\cos(t'\,u_{\rm h}^{(j)})\right.\\ \nonumber
&&\ -\left.4\,g^2\,\int_0^\infty\,dt'\,\hat{K}(2gt,2gt')\,\oldtildesigma(t')\right].
\end{eqnarray}
The corresponding anomalous dimension can be easily shown to be given by
\begin{eqnarray}\label{adwithholes}
\gamma=8\,g^2\,\log{M}\,&&\left(1-\frac{1}{\log M}\,\sum_{j=3}^{L}\,\int_0^\infty\,dt\,\frac{J_1(2gt)}{2gt}e^{-t/2}\,\cos(tu_{\rm h}^{(j)}) \right.\\ \nonumber
&&\ \ -\left. 8\,\int_0^\infty\,dt\,\frac{J_1(2gt)}{2gt}\,
\oldtildesigma(t)\right).
\end{eqnarray}
The distribution of the small holes is found from
\begin{equation}\label{holesdef}
Z(u_{\rm h}^{j})=\pi\,(2\,n_{\rm h} ^j+\delta-1),
\end{equation}
which in Fourier space reads
\begin{equation} \label{holesequation}
\frac{i}{\pi} \,\int^{\infty}_{0}\, \sin \left(t\,u_{\rm h}^j \right )\,\hat{Z}(t) = \pi\,(2\,n_{\rm h} ^j+\delta-1).
\end{equation}
Plugging \eqref{stripoffFT} into \eqref{holesequation} and observing that (see section \ref{sec:aexp})
\begin{eqnarray}\label{univ}
\nonumber F'(x,y)&\equiv&\int^{\infty}_0 dt\, \cos{t x} \, \frac{e^{\frac{t}{2}}-\cos{t\, y}}{e^t-1}\\
\nonumber &=& \frac{1}{4}\,\bigg(\psi \left(i\,(x-y)\right)+\psi \left(-i\,(x-y)\right)+\psi \left(i\,(x+y) \right)+\psi \left(-i\,(x+y) \right) \\
&&\quad \ -2\,\psi\left(\frac{1}{2}-i\,x \right)-2\,\psi \left(\frac{1}{2}+i\,x \right) \bigg),
\end{eqnarray}
one easily derives from \eqref{holesequation}
\begin{eqnarray}\label{smallholes}
2\,\pi\,n_{\rm h}^{(k)}&=&4\,F(u_{\rm h}^{(k)},u_{\rm h}^{(1)})-16\,\log M\,\int_0^\infty\,dt\,\frac{\oldtildesigma(t)}{t}\,e^{t/2}\,\sin(tu_{\rm h}^{(k)}).
\end{eqnarray}
Introducing the density of holes $\holerho(u)$ it follows from 
\eqref{smallholes} that
\begin{equation}\label{holessigmahl}
\LoverlogM\,\holerho(u)=\frac{2}{\pi\,\log M}\,F'(u,u_{\rm h}^{(1)})-\frac{8}{\pi}\,\int_0^\infty\,dt\,\oldtildesigma(t)\,e^{t/2}\,\cos(t\,u).
\end{equation}
Note that $\frac{2}{\pi}\,\frac{1}{M}\,F'(u,u_{\rm h}^{(1)})$ is at large values of $M$ essentially the Korchemsky density $\rho_0(u)$,
i.e.~\eqref{1loopdenssol} after scaling back 
$u=M\,\bar u$, $\rho_0(u)=1/M\,\bar \rho_0(\bar u)$,
up to small corrections at the boundaries of the distribution of the roots. Since the small holes occupy a finite interval $(-a,a)$ one can safely take the large $M$ limit\footnote{%
The magnon density is related to $\oldtildesigma(t)$ by a similar formula
\begin{equation}\label{magnonssigma}\rho_m(u)=\frac{2}{\pi}\,\frac{1}{M}\,F'(u,u_{\rm h}^{(1)})-\frac{8\,\log{M}}{\pi\,M}\,\int_0^\infty\,dt\,\oldtildesigma(t)\,e^{t/2}\,\cos(t\,u).
\nonumber \end{equation}
}%
\begin{equation}\label{expforholes}
F'(u,u_{\rm h}^{(1)})=\log M + \Op(M^0) \qquad \qquad u \in (-a,a).
\end{equation}
After replacing the sum in \eqref{densityAM} by an integral and using the above density we find
\begin{eqnarray}\label{sigmatilde}
\nonumber\oldtildesigma(t)&=&\frac{t}{e^t-1}\left[-\frac{\LoverlogM}{8\,t}\,J_0(2gt)+\hat{K}_{\rm h}(t,0;a)-4\,\int_0^\infty dt'\,\hat{K}_{\rm h}(t,t';a)\oldtildesigma(t')\right.\\ \nonumber
&&\left.\quad\,\,\qquad+g^2\,\hat{K}(2gt,0)-4\,g^2\,\int_0^\infty dt'\,\hat{K}(2gt,2gt')\,\oldtildesigma(t') \right.\\ \nonumber
\lefteqn{\left.~~~
- 4\,g^2\,\int_0^\infty dt'\,t'\,\hat{K}(2gt,2gt')\,\bigg(\hat{K}_{\rm h}(t',0;a)
-4\,\int_0^\infty dt''\,\hat{K}_{\rm h}(t',t'')\,\oldtildesigma(t'')\bigg)\right]}\\
\end{eqnarray}
where $\hat{K}_{{\rm h}}(t,t';a)$ is the one-loop kernel given in 
\eqref{ktilde}. 
The endpoints can be obtained from the normalization condition
\begin{equation}\label{endpoints}
\int_{-a}^{a} du\,\rho(u)=1
\end{equation}
which implies \eqref{normalizationtspace}. 
Inserting \eqref{normalizationtspace} into \eqref{sigmatilde}
we find the final integral equation \eqref{besartige} anounced in
the introduction. Likewise, the anomalous dimension 
\eqref{adwithholes} may be reexpressed and simplified as
\begin{eqnarray}\label{energy1}
\nonumber \gamma&=&8\,g^2\,\log M\,\left[1-8\,\int_0^\infty\,dt\,\frac{J_1(2gt)}{2gt}\,t\,\hat{K}_{\rm h}(t,0;a)\right.\\
\nonumber&-&8\,\left.\int_0^\infty\,dt\,\frac{J_1(2gt)}{2gt}\,\left(\oldtildesigma(t)-4\,t\,\int_0^\infty\,dt'\,\hat{K}_{\rm h}(t,t';a)\,\oldtildesigma(t')\right)\right]\\
&=&16\,\log M\,\left(\oldtildesigma(0)+\frac{\LoverlogM}{16}\right).
\end{eqnarray}
%
%
%
%
%
This concludes our derivation of the equations determining the
generalized scaling function $f(g,\LoverlogM)$ in \eqref{newlog}.
Let us now apply them to obtain the first few terms in the
double expansion of this function in powers of $g$ and $\LoverlogM$.

\subsection{Weak Coupling Expansion}

The equation \eqref{sigmatilde} is solved iteratively with relative ease
in a double-perturbative series in $g$ and the gap parameter
$a$. As in the one-loop case in section \ref{sec:1-loop}
one then inverts \eqref{normalizationtspace} to obtain
$a(\LoverlogM$) as a power series in $\LoverlogM$. 
This then yields the fluctuation density $\oldtildesigma(t)$ as
a series in $g$ and $\LoverlogM$. It starts out as
\begin{eqnarray}\label{weak2}
\nonumber\oldtildesigma(t)&=&g^2\,\oldtildesigma_{\mbox{\tiny BES}}(t)\,\\
\nonumber&+&\,\LoverlogM\left(-\frac{1}{8}\frac{1}{e^{t/2}\,+\,e^t}\,+\,\frac{g^2}{8}\frac{t\,(t\,-\,4\log 2)}{e^t\,-\,1}\right.\\
\nonumber& &\,\left.
~~~~+g^4\frac{t}{e^t\,-\,1}\frac{1}{96}\left(-\,3t^3\,-\,4\pi^2t\,+\,16\pi^2\log 2\,+\,24t^2\log 2\,+\,96\zeta(3)\right)\,+\,\dots\right)\\
\nonumber&+&\,\LoverlogM^2\,\times 0\\
\nonumber&+&\,j^3
\left(
-\,\frac{\pi^2}{1536}t^2e^{-t}\mbox{csch}(t/2)\,+\,\frac{g^2\pi^2}{384}\frac{t\,(14\zeta(3)\,-\,\pi^2te^{-t/2})}{e^t\,-\,1}
\right.\\\nonumber& &\,%
\left.%
~~~~~~+\frac{g^4\pi^2}{2304}\frac{t}{e^t\,-\,1} 
\big(
3\pi^4t\,-\,\pi^4te^{-t/2}\,
\right.\\\nonumber& &\,%
\left.%
~~~~~~~~~~~~~~~~~~~~~~~~~~~~~~+\,140\pi^2\zeta(3)
-\,42\zeta(3)t^2\,-\,2232\zeta(5)
\big)\,
+\,\dots
\right.\bigg)\\
&+&\dots
\end{eqnarray}
The generalized scaling function at weak coupling is simply given 
via \eqref{genscalingfunction} by evaluating the fluctuation density 
at $t=0$. Let us define an infinite set of functions $\{f_n(g)\}$ as
\begin{eqnarray}\label{f-family-repeat}
f(g,\LoverlogM)=f(g)+\sum_{n=1}^\infty f_n(g)\,\LoverlogM^n\, .
\end{eqnarray}
The first one $f_1(g)$ is
\begin{eqnarray}\label{f1}
\nonumber f_1(g)&=&-8\,g^2\,\log 2+g^4\left(\frac{8}{3}\,\pi^2\,\log2+16\,\zeta(3)\right)
-g^6\bigg(\frac{88}{45}\,\pi^4\,\log 2+\frac{8}{3}\,\pi^2\,\zeta(3)
+168\,\zeta(5)\bigg)\\
&+&g^8\left(\frac{584}{315}\,\pi^6\,\log 2+\frac{8}{5}\,\pi^4\,\zeta(3)+64\,\log 2\,\zeta(3)^2+\frac{88}{3}\,\pi^2\,\zeta(5)+1840\,\zeta(7)\right)+ \ldots\nonumber \\
\end{eqnarray}
Note that $f_1(g)$ is special as it can be obtained from 
\eqref{subleading} by keeping only terms proportional to $L$. To this order the hole momenta are set to zero. Only at orders higher than
linear in $\LoverlogM$ 
one needs to take into account the ``dynamics'' of the holes. We then
find for $f_1(g), \ldots f_4(g)$ 
\begin{eqnarray}\label{f1alt}
\nonumber f_1(g)&=&- f(g)\,\log 2+16\,g^4\,\zeta(3)-g^6\bigg(\frac{8}{3}\,\pi^2\,\zeta(3)+168\,\zeta(5)\bigg)\\
&+&g^8\left(\frac{8}{5}\,\pi^4\,\zeta(3)+\frac{88}{3}\,\pi^2\,\zeta(5)+1840\,\zeta(7)\right) +\ldots \nonumber \\  \\ \nonumber \\
\label{f2}%
f_2(g)&=&0 \\ \nonumber \\
\nonumber f_3(g)&=&\frac{7}{12}\,g^2\,\pi^2\,\zeta(3)+g^4\left(\frac{35}{36}\,\pi^4\,\zeta(3)-\frac{31}{2}\,\pi^2\,\zeta(5)\right)\\
\nonumber&+&g^6\left(-\frac{73}{540}\,\pi^6\,\zeta(3)-\frac{155}{6}\,\pi^4\,\zeta(5)+\frac{635}{2}\,\pi^2\,\zeta(7)\right)\\
&+&g^8\,\left(\frac{7}{108}\,\pi^8\,\zeta(3)+\frac{182}{3}\,\pi^2\,\zeta(3)^3+\frac{28}{15}\,\pi^6\,\zeta(5)+\frac{3175}{6}\,\pi^4\,\zeta(7)-\frac{17885}{3}\,\pi^2\,\zeta(9)\right)\nonumber \\
&+&\label{f3}%
\ldots \end{eqnarray}
\begin{eqnarray}
\nonumber
f_4(g)&=&-\frac{7}{6}\,g^2\,\pi^2\,\log 2 \,\zeta(3)+g^4\left(-\frac{77}{18}\,\pi^4\,\log 2\,\zeta(3)+\frac{49}{6}\,\pi^2\,\zeta(3)^2+31\,\pi^2\,\log 2\,\zeta(5)\right)\\
\nonumber&+&g^6\,\left(-\frac{767}{270}\,\pi^6\,\log 2\,\zeta(3)+\frac{385}{18}\,\pi^4\,\zeta(3)^2+\frac{341}{3}\,\pi^4\,\log 2\,\zeta(5)\right.\\
\nonumber&&\left.  \quad \ \ \,-\frac{651}{2}\,\pi^2\,\zeta(3)\,\zeta(5)-635\,\pi^2\,\log2\,\zeta(7)\right)\\
\nonumber&+&g^8\left(\frac{307}{270}\,\pi^8\,\log 2\,\zeta(3)+\frac{91}{15}\,\pi^6\,\zeta(3)^2-252\,\pi^2\,\log 2\,\zeta(3)^3+\frac{1184}{15}\,\pi^6\,\log 2\,\zeta(5)\right.\\
\nonumber
&&\left.\quad \ \ \,-\frac{15011}{18}\,\pi^4\,\zeta(3)\,\zeta(5)+2883\,\pi^2\,\zeta(5)^2-\frac{6985}{3}\,\pi^4\,\log 2\,\zeta(7)\right.\\
&&\left. \quad \ \ +\frac{17780}{3}\,\pi^2\,\zeta(3)\,\zeta(7)+\frac{35770}{3}\,\pi^2\,\log 2\,\zeta(9)\right) \nonumber\\ 
\label{f4}%
&+&\ldots \,. 
\end{eqnarray}
At fixed $j$, we observe a constant degree of transcendentality 
\cite{Kotikov:2002ab}
of all terms contributing to a given order of perturbation theory in the 
coupling g. Interestingly, the converse is not true,
as may already be seen from the one-loop result \eqref{ellexpansion}.

As was announced earlier the function $f_2(g)$ is identically
zero, indicating that all terms of order $\LoverlogM^2$ in the
$\LoverlogM$-expansion of $f(g,\LoverlogM)$ are absent to 
all orders in the coupling constant $g$. This is easily proven
directly from our equations.
Some potentially related very interesting observations at 
strong coupling were made in \cite{Radu+Arkady}.
Roiban and Tseytlin found some intriguing evidence that terms of the
form $\LoverlogM^2\, \log^k \LoverlogM$ might upon
resummation indeed result in a vanishing $\LoverlogM^2$
contribution, {\it cf} also the discussion in the introduction.


\subsection*{Acknowledgments}
We would like to thank Vladimir Bazhanov, Andrei Belitsky,
Nick Dorey, Volodya Kazakov, Gregory Korchemsky,
Lev Lipatov,  Pedro Vieira, Stefan Zieme
and, especially, Radu Roiban and Arkady Tseytlin
for useful discussions.
M.S.~thanks the INI Cambridge and the organizers of the
workshop {\it Strong Fields, Integrability and Strings},
N.~Dorey, S.~Hands and N.~MacKay, for hospitality while
working on parts of this manuscript.
A.~Rej thanks the INI Cambridge for hospitality during the
finishing phase of the project.

\appendix
\section{The Non-Linear Term}
\label{app:nonlinearterm}

In this appendix we will discuss the integrals involving the non-linear term. For simplicity we will confine ourselves to the one-loop case, where it is sufficient to consider
\begin{equation} \label{imlog}
{\cal I }(u)=\lim_{\alpha \to \infty}\int_{-\alpha}^\alpha \frac{dv}{\pi}\,i\,\frac{d}{du}\log\frac{\Gamma(-i\,(u-v))}{\Gamma(i\,(u-v))}\,\mbox{Im}\log\left[1+(-1)^\delta\, e^{i\,Z(v+i\,0)}\right]
\end{equation}
We first note that the function
\begin{equation}
{\cal L}(u)=\mbox{Im}\log\left[1+(-1)^\delta\, e^{i\,Z(u+i\,0)}\right]
\end{equation}
is smooth apart from a finite numbers of points, namely when $u$ is equal to the magnon or the hole rapidity. A closer inspection reveals that
\begin{equation}
{\cal L}(u_i-\epsilon)=\pi \qquad \qquad {\cal L}(u_i+ \epsilon)=-\pi
\end{equation}
where $u_i$ denotes either a hole or a magnon rapidity. We will assume that the small holes and the magnons are densely distributed along the real axis, as this is the case for the limits discussed in this paper. It is easy to convince oneself that the integral 
$\eqref{imlog}$ gets the dominant contribution from $(-\alpha,-\frac{M}{2})\cup(\frac{M}{2},\alpha)$. Because the small roots and magnons are, at large values of $M$, densely and symmetrically distributed on $(-\frac{M}{2},\frac{M}{2})$ this part of the integral contributes starting at $\Op \left(\frac{1}{M^2}\right)$ only. Assuming $v\in (-\alpha,-\frac{M}{2})\cup(\frac{M}{2},\alpha)$ we may expand the integrand in a power series in $u$. Because of the antisymmetry of the counting function only odd powers of $u$ survive the integration. Thus we may write
\begin{eqnarray}\label{integrandexpansion}
&&\nonumber i\,\frac{d}{du}\log\frac{\Gamma(-i\,(u-v))}{\Gamma(i\,(u-v))}=i\,\bigg(\psi_1 (-i\,v)-\psi_1(i\,v)\bigg)u-\frac{i}{6}\,\bigg(\psi_3 (-i\,v)-\psi_3(i\,v)\bigg)u^3\\
&&\qquad \qquad \qquad \qquad \quad \ \ +\ \Op(u^5)+\mbox{even terms in}~v\, .
\end{eqnarray}
On the other hand from the definition of the counting function
we have
\begin{equation}\label{Imloglargev}
{\cal L}(v)=-\frac{L+2M}{2\,v}+O\bigg(\frac{1}{v^3}\bigg) \qquad \qquad v > \frac{M}{\sqrt{2}}\, .
\end{equation}
Plugging \eqref{integrandexpansion} and \eqref{Imloglargev} into \eqref{imlog} we find
\begin{equation}\label{twologtwou}
{\cal I}(u)=\xi\,u+\Op\left(\frac{u^3}{M^2}\right).
\end{equation}
To fix the constant $\xi$ it is necessary to extend the expansion in \eqref{Imloglargev} to the whole interval $v\in (\frac{M}{2},\infty)$. However there is a much simpler method. Since the above discussion is not sensitive to the value of $L$ we may set $L=2$. Then we may compute the corresponding anomalous dimension plugging \eqref{Zuspace} together with \eqref{imlog} into \eqref{cruciali}. Comparison with the exact one-loop result $\gamma_1=8\,S_1(M)$ fixes $\xi$ to be
\begin{equation}
\xi=2\,\log{2}
\end{equation}
Numerically we have checked that the expansion \eqref{twologtwou} breaks only around the small neighborhood of $\pm \frac{M}{2}$. This suggests that the radius of convergence of \eqref{twologtwou} lies closely to the edge of the magnon distribution.




\begin{thebibliography}{99}

\bibitem{Collins:1989bt}
 J.~C.~Collins,
 {\it ``Sudakov form factors,''}
 Adv.\ Ser.\ Direct.\ High Energy Phys.\  {\bf 5} (1989) 573,
 {\tt hep-ph/0312336}.

\bibitem{Korchemsky:1988si}
 G.~P.~Korchemsky,
 {\it ``Asymptotics of the Altarelli-Parisi-Lipatov Evolution Kernels of Parton Distributions,''}
 Mod.\ Phys.\ Lett.\  A {\bf 4} (1989) 1257.
$\bullet$
 G.~P.~Korchemsky and G.~Marchesini,
 {\it ``Structure function for large x and renormalization of Wilson loop,''}
 Nucl.\ Phys.\  B {\bf 406} (1993) 225,
 {\tt hep-ph/9210281}.

\bibitem{Belitsky:2006en}
 A.~V.~Belitsky, A.~S.~Gorsky and G.~P.~Korchemsky,
 {\it ``Logarithmic scaling in gauge / string correspondence,''}
 Nucl.\ Phys.\  B {\bf 748} (2006) 24,
 {\tt hep-th/0601112}.

\bibitem{Beisert:2003tq}
 N.~Beisert, C.~Kristjansen and M.~Staudacher,
 {\it ``The dilatation operator of $\mathcal{N} = 4$ super Yang-Mills theory,''}
 Nucl.\ Phys.\  B {\bf 664} (2003) 131,
 {\tt hep-th/0303060}.

\bibitem{Beisert:2005fw}
N.~Beisert and M.~Staudacher,
{\it ``Long-range $\alg{psu}(2,2|4)$ Bethe ans\"atze for gauge theory
and strings,''}
Nucl.\ Phys.\ B {\bf 727} (2005) 1,
{\tt hep-th/0504190}.

\bibitem{Eden:2006rx}
 B.~Eden and M.~Staudacher,
 {\it ``Integrability and transcendentality,''}
 J.\ Stat.\ Mech.\  {\bf 0611} (2006) P014,
 {\tt hep-th/0603157}.

\bibitem{Beisert:2006ez}
 N.~Beisert, B.~Eden and M.~Staudacher,
 {\it ``Transcendentality and crossing,''}
 J.\ Stat.\ Mech.\  {\bf 0701} (2007) P021,
 {\tt hep-th/0610251}.

\bibitem{Bern:2006ew}
 Z.~Bern, M.~Czakon, L.~J.~Dixon, D.~A.~Kosower and V.~A.~Smirnov,
 {\it ``The Four-Loop Planar Amplitude and Cusp Anomalous Dimension in Maximally Supersymmetric Yang-Mills Theory,''}
 Phys.\ Rev.\  D {\bf 75} (2007) 085010,
 {\tt hep-th/0610248}.
$\bullet$
 F.~Cachazo, M.~Spradlin and A.~Volovich,
 {\it ``Four-Loop Cusp Anomalous Dimension From Obstructions,''}
 Phys.\ Rev.\  D {\bf 75} (2007) 10501,
 {\tt hep-th/0612309}.

\bibitem{Bern:2007ct}
 Z.~Bern, J.~J.~M.~Carrasco, H.~Johansson and D.~A.~Kosower,
 {\it ``Maximally supersymmetric planar Yang-Mills amplitudes at five loops,''}
 {\tt 0705.1864 [hep-th]}.

\bibitem{Kotikov:2006ts}
 A.~V.~Kotikov and L.~N.~Lipatov,
 {\it``On the highest transcendentality in $\mathcal{N}= 4$ SUSY'',}
 Nucl.\ Phys.\  B {\bf 769} (2007) 217,
 {\tt hep-th/0611204}.

\bibitem{Benna:2006nd}
 M.~K.~Benna, S.~Benvenuti, I.~R.~Klebanov and A.~Scardicchio,
 {\it ``A test of the AdS/CFT correspondence using high-spin operators,''}
 Phys.\ Rev.\ Lett.\  {\bf 98} (2007) 131603,
 {\tt hep-th/0611135}.
$\bullet$
 L.~F.~Alday, G.~Arutyunov, M.~K.~Benna, B.~Eden and I.~R.~Klebanov,
 {\it ``On the Strong Coupling Scaling Dimension of High Spin Operators,''}
JHEP {\bf 0704} (2007) 082,
{\tt hep-th/0702028}.
$\bullet$
 I.~Kostov, D.~Serban and D.~Volin,
 {\it ``Strong coupling limit of Bethe ansatz equations,''}
 Nucl.\ Phys.\  B {\bf 789} (2008) 413,
 {\tt hep-th/0703031}.
$\bullet$
 M.~Beccaria, G.~F.~De Angelis and V.~Forini,
 {\it ``The scaling function at strong coupling from the quantum
 string Bethe equations,''}
 JHEP {\bf 0704} (2007) 066,
 {\tt hep-th/0703131}.

\bibitem{Basso:2007wd}
 B.~Basso, G.~P.~Korchemsky and J.~Kota\'nski,
 {\it ``Cusp anomalous dimension in maximally supersymmetric Yang-Mills theory at strong coupling,''}
 Phys.\ Rev.\ Lett.\  {\bf 100} (2008) 091601,
 {\tt 0708.3933 [hep-th]}.

\bibitem{Eden}
B.~Eden, unpublished,
Talk at the 12th Claude Itzykson Meeting:\\
{\it ``Integrability in Gauge and String Theory''},
18-22 June 2007, Paris, France,\\
{\tt http://www-spht.cea.fr/Meetings/Rencitz2007/eden.pdf}.

\bibitem{Gubser:2002tv}
 S.~S.~Gubser, I.~R.~Klebanov and A.~M.~Polyakov,
 {\it``A semi-classical limit of the gauge/string correspondence'',}
 Nucl.\ Phys.\ B {\bf 636} (2002) 99,
 {\tt hep-th/0204051}.

\bibitem{Frolov:2002av}
S.~Frolov and A.~A.~Tseytlin,
{\it ``Semiclassical quantization of rotating
superstring in $AdS_5 \times S^5$,''}
JHEP {\bf 0206} (2002) 007,
{\tt hep-th/0204226}.

\bibitem{Beisert:2003ea}
 N.~Beisert, S.~Frolov, M.~Staudacher and A.~A.~Tseytlin,
 {\it ``Precision spectroscopy of AdS/CFT,''}
 JHEP {\bf 0310} (2003) 037,
 {\tt hep-th/0308117}.

\bibitem{Roiban:2007jf}
 R.~Roiban, A.~Tirziu and A.~A.~Tseytlin,
 {\it ``Two-loop world-sheet corrections in $AdS_5 \times S^5$ superstring,''}
 JHEP {\bf 0707} (2007) 056,
 {\tt 0704.3638 [hep-th]}.

\bibitem{Roiban:2007dq}
 R.~Roiban and A.~A.~Tseytlin,
 {\it ``Strong-coupling expansion of cusp anomaly from quantum superstring,''}
 JHEP {\bf 0711} (2007) 016,
 {\tt 0709.0681 [hep-th]}.

\bibitem{Kotikov:2007cy}
 A.~V.~Kotikov, L.~N.~Lipatov, A.~Rej, M.~Staudacher and V.~N.~Velizhanin,
 {\it ``Dressing and Wrapping,''}
 J.\ Stat.\ Mech.\  {\bf 0710} (2007) P10003,
 {\tt 0704.3586 [hep-th]}.

\bibitem{Frolov:2006qe}
 S.~Frolov, A.~Tirziu and A.~A.~Tseytlin,
 {\it ``Logarithmic corrections to higher twist scaling at strong coupling from AdS/CFT,''}
 Nucl.\ Phys.\  B {\bf 766} (2007) 232,
 {\tt hep-th/0611269}.

\bibitem{Alday:2007mf}
 L.~F.~Alday and J.~M.~Maldacena,
 {\it ``Comments on operators with large spin,''}
 JHEP {\bf 0711} (2007) 019
 {\tt 0708.0672 [hep-th]}.

\bibitem{Rej:2007vm}
 A.~Rej, M.~Staudacher and S.~Zieme,
 {\it ``Nesting and dressing,''}
 J.\ Stat.\ Mech.\  {\bf 0708} (2007) P08006,
 {\tt hep-th/0702151}.

\bibitem{Radu+Arkady}
R.~Roiban and A.A.~Tseytlin,
{\it ``Spinning superstrings at 2-loops: strong-coupling corrections 
to dimensions of large-twist SYM operators''},
Phys.\ Rev.\  D {\bf 77} (2008) 066006,
{\tt 0712.2479 [hep-th]}.

\bibitem{Casteill:2007ct}
 P.~Y.~Casteill and C.~Kristjansen,
 {\it ``The Strong Coupling Limit of the Scaling Function from the Quantum String Bethe Ansatz,''}
 Nucl.\ Phys.\  B {\bf 785} (2007) 1,
 {\tt 0705.0890 [hep-th]}.

\bibitem{Arutyunov:2004vx}
G.~Arutyunov, S.~Frolov and M.~Staudacher, 
{\it ``Bethe ansatz for quantum strings,''} 
JHEP {\bf 0410} (2004) 016, 
{\tt hep-th/0406256}.

\bibitem{Bena:2003wd}
 I.~Bena, J.~Polchinski and R.~Roiban,
 {\it ``Hidden symmetries of the $AdS_5 \times S^5$ superstring,''}
 Phys.\ Rev.\  D {\bf 69} (2004) 046002,
 {\tt hep-th/0305116}.

\bibitem{Kazakov:2004qf}
 V.~A.~Kazakov, A.~Marshakov, J.~A.~Minahan and K.~Zarembo,
 {\it ``Classical / quantum integrability in AdS/CFT,''}
 JHEP {\bf 0405} (2004) 024,
 {\tt hep-th/0402207}.
$\bullet$ 
 N.~Beisert, V.~A.~Kazakov, K.~Sakai and K.~Zarembo,
 {\it ``The algebraic curve of classical superstrings on $AdS_5 \times S^5$,''}
 Commun.\ Math.\ Phys.\  {\bf 263} (2006) 659,
 {\tt hep-th/0502226}.

\bibitem{Hernandez:2006tk}
 R.~Hern\'andez and E.~L\'opez,
 {\it ``Quantum corrections to the string Bethe ansatz,''}
 JHEP {\bf 0607} (2006) 004,
 {\tt hep-th/0603204}.

\bibitem{Freyhult:2006vr}
 L.~Freyhult and C.~Kristjansen,
 {\it ``A universality test of the quantum string Bethe ansatz,''}
 Phys.\ Lett.\  B {\bf 638} (2006) 258,
 {\tt hep-th/0604069}.

\bibitem{Gromov:2007ky}
 N.~Gromov and P.~Vieira,
 {\it ``Complete 1-loop test of AdS/CFT,''}
 {\tt 0709.3487 [hep-th]}.

\bibitem{Dorey:2006zj}
 N.~Dorey and B.~Vicedo,
 {\it ``On the dynamics of finite-gap solutions in classical string theory,''}
 JHEP {\bf 0607} (2006) 014,
 {\tt hep-th/0601194}.

\bibitem{Feverati:2006tg}
 G.~Feverati, D.~Fioravanti, P.~Grinza and M.~Rossi,
 {\it ``On the finite size corrections of anti-ferromagnetic anomalous
   dimensions in $\mathcal{N} = 4$ SYM,''}
 JHEP {\bf 0605} (2006) 068,
 {\tt hep-th/0602189}.
$\bullet$
 G.~Feverati, D.~Fioravanti, P.~Grinza and M.~Rossi,
 {\it ``Hubbard's adventures in $\mathcal{N} = 4$ SYM-land? Some non-perturbative considerations on finite length operators,''}
 J.\ Stat.\ Mech.\  {\bf 0702} (2007) P001,
 {\tt hep-th/0611186}.
$\bullet$
 D.~Fioravanti and M.~Rossi,
 {\it ``On the commuting charges for the highest dimension SU(2) operator in planar ${\cal N}=4$ SYM,''}
 JHEP {\bf 0708} (2007) 089,
 {\tt 0706.3936 [hep-th]}.
$\bullet$  
 D.~Bombardelli, D.~Fioravanti and M.~Rossi,
 {\it ``Non-linear integral equations in ${\cal {N}}=4$ SYM,''}
 {\tt 0711.2934 [hep-th]}.

\bibitem{Lipatov:1997vu}
 L.~N.~Lipatov,
  {\it "Evolution equations in QCD",}
    in ``Perspectives in Hadronic Physics,''
    Proceedings of the Conference, ICTP, Trieste, Italy, 12-16 May 1997,
    eds.~S.~Boffi, C.~Ciofi Degli Atti and M.~Giannini,
    World Scientific (Singapore, 1998).

\bibitem{Braun:1998id}
 V.~M.~Braun, S.~E.~Derkachov and A.~N.~Manashov,
 {\it ``Integrability of three-particle evolution equations in {QCD},''}
 Phys.\ Rev.\ Lett.\  {\bf 81} (1998) 2020,
 {\tt hep-ph/9805225}.
$\bullet$
 V.~M.~Braun, S.~E.~Derkachov, G.~P.~Korchemsky and A.~N.~Manashov,
 {\it ``Baryon distribution amplitudes in {QCD},''}
 Nucl.\ Phys.\ B {\bf 553} (1999) 355,
 {\tt hep-ph/9902375}.
$\bullet$
 A.~V.~Belitsky,
 {\it ``Fine structure of spectrum of twist-three operators in {QCD},''}
 Phys.\ Lett.\ B {\bf 453} (1999) 59,
 {\tt hep-ph/9902361}.
$\bullet$
 A.~V.~Belitsky,
 {\it ``Renormalization of twist-three operators and integrable lattice models,''}
 Nucl.\ Phys.\ B {\bf 574} (2000) 407,
 {\tt hep-ph/9907420}.

\bibitem{Minahan:2002ve}
 J.~A.~Minahan and K.~Zarembo,
 {\it ``The Bethe ansatz for $\mathcal{N}= 4$ super Yang-Mills'',}
 JHEP {\bf 0303} (2003) 013,
 {\tt arXiv:hep-th/0212208}.

\bibitem{Beisert:2003yb}
 N.~Beisert and M.~Staudacher,
 {\it``The $\mathcal{N}= 4$ SYM integrable super spin chain'',}
 Nucl.\ Phys.\ B {\bf 670}, 439 (2003),
 {\tt arXiv:hep-th/0307042}.

\bibitem{Korchemsky:1995be}
G.~P.~Korchemsky,
{\it ``Quasiclassical QCD pomeron,''}
Nucl.\ Phys.\ B {\bf 462} (1996) 333,
{\tt hep-th/9508025}.

\bibitem{Staudacher:2004tk}
 M.~Staudacher,
 {\it ``The factorized S-matrix of CFT/AdS'',}
 JHEP {\bf 0505}, 054 (2005),
 {\tt hep-th/0412188}.
$\bullet$
 N.~Beisert,
 {\it ``The $\mathfrak{su}(2|2)$ dynamic S-matrix,''}
 {\tt hep-th/0511082}.

\bibitem{Kotikov:2002ab}
 A.~V.~Kotikov and L.~N.~Lipatov,
 {\it ``DGLAP and BFKL equations in the $\superN = 4$ supersymmetric gauge theory,''}
 Nucl.\ Phys.\  B {\bf 661} (2003) 19
 [Erratum-ibid.\  B {\bf 685} (2004) 405],
 {\tt hep-ph/0208220}.

\end{thebibliography}
\end{document}